\pdfoutput=1 
\newif\ifarXiv
\arXivtrue 

\ifarXiv
\documentclass[aps,pre,reprint,superscriptaddress]{revtex4-1}

\else

\documentclass[NETN]{stjour}

\articletype{Research}

\supportinginfo{}

\received{}
\accepted{}
\published{}
\handlingeditor{}

\setdoi{}

\fi

\newcommand{\nn}{\nonumber}
\newcommand{\ie}{{i.e.},~}
\newcommand{\eg}{{e.g.}~}

\newcommand{\fig}[1]{Figure~\ref{fig:#1}}

\newcommand{\secRef}[1]{Section~\ref{sec:#1}}
\newcommand{\appRef}[1]{Appendix~\ref{app:#1}}
\newcommand{\eq}[1]{Eq.~(\ref{eq:#1})}

\usepackage{microtype}
\usepackage{amsmath,amssymb,bm,siunitx,mathtools}
\usepackage{graphicx}
\graphicspath{{images/}}
\usepackage{bm}
\usepackage{enumitem}
\usepackage{lineno}

\newcommand{\paperTitle}{Inferring network properties from time series using transfer entropy and mutual information: validation of multivariate versus bivariate approaches}
\newcommand{\shortTitle}{Inferring network properties from time series}
\newcommand{\theKeywords}{directed connectivity, functional connectivity, network inference, multivariate transfer entropy, information theory, complex networks}

\begin{document}

\ifarXiv

\title{\paperTitle}


\author{Leonardo Novelli}
\email[]{leonardo.novelli@sydney.edu.au}
\affiliation{Centre for Complex Systems, Faculty of Engineering, The University of Sydney, Sydney, Australia}
\author{Joseph T. Lizier}
\affiliation{Centre for Complex Systems, Faculty of Engineering, The University of Sydney, Sydney, Australia}


\date{\today}

\fi

\begin{abstract}
Functional and effective networks inferred from time series are at the core of network neuroscience.
Interpreting their properties requires inferred network models to reflect key underlying structural features; however, even a few spurious links can distort network measures, challenging functional connectomes.
We study the extent to which micro- and macroscopic properties of underlying networks can be inferred by algorithms based on mutual information and bivariate/multivariate transfer entropy.
The validation is performed on two macaque connectomes and on synthetic networks with various topologies (regular lattice, small-world, random, scale-free, modular).
Simulations are based on a neural mass model and on autoregressive dynamics (employing Gaussian estimators for direct comparison to functional connectivity and Granger causality).
We find that multivariate transfer entropy captures key properties of all networks for longer time series.
Bivariate methods can achieve higher recall (sensitivity) for shorter time series but are unable to control false positives (lower specificity) as available data increases.
This leads to overestimated clustering, small-world, and rich-club coefficients, underestimated shortest path lengths and hub centrality, and fattened degree distribution tails.
Caution should therefore be used when interpreting network properties of functional connectomes obtained via correlation or pairwise statistical dependence measures, rather than more holistic (yet data-hungry) multivariate models.
\end{abstract}

\ifarXiv
\maketitle
\fi

\section{Introduction}
Functional and effective network inference in neuroscience typically involves pre-processing the data, defining the parcellation, extracting the time series, inferring the links in the model network, and measuring network properties, \eg to compare patients and controls or to predict phenotype~\citep{Bassett2017,Fornito2016}.
Each step in the pipeline requires making modelling and analysis choices, whose influence on the final results is the subject of ongoing research~\citep{Zalesky2010,Zalesky2016,Aquino2020,Cliff2020}.
As part of this effort, we study how the choice of different inference algorithms affects the properties of the resulting network model in comparison to the underlying structural network, and whether the ability to accurately reflect these properties changes across different underlying structural networks.

The structure (or topology) of a network can be described at multiple scales~\citep{Bassett2017}: from the microscopic (individual links), to the mesoscopic (modules and motifs) and the macroscopic (summary statistics, such as average shortest-path length and measures of small-worldness)~\citep{Rubinov2010}.
At each scale, the structure is associated with development, ageing, cognition, and neuropsychiatric diseases~\citep{Xia2020}.
Previous studies have assessed the performance of different network inference algorithms in identifying the structural links at the microscale~\citep{Novelli2019,Runge2018arXiv,Sun2015,Kim2016,Razi2015}.
The goal of this work is to extend the assessment to all scales and to a variety of topologies, across a range of related network inference algorithms.
We link the performance at the microscale to the resulting network properties at the macroscale, and describe how this changes as a function of the overall topology.

We compare bivariate and multivariate approaches for inferring network models, employing statistical dependence measures based on information theory~\citep{Shannon1948}.
These approaches include \textit{functional network inference}, which produces models of networks of pairwise or bivariate statistical relationships between nodes, and can either quantify undirected statistical dependence, in the case of mutual information (MI)~\citep{Cover2005}, or directed dependence, in the case of transfer entropy (TE)~\citep{Schreiber2000,Bossomaier2016}.
These approaches also include \textit{effective network inference}, which is intended to produce the simplest possible circuit models that explain the observed responses~\citep{Aertsen1989}.
In this class, we evaluate the use of multivariate TE, which, in contrast to the bivariate approaches, aims to minimise spurious links and infer minimal models of the parent \textit{sets} for each target node in the network.

All of these inference techniques seek to infer a network \textit{model} of the relationships between the nodes in a system.
Different methods capture different aspects of these relationships and don't necessarily seek to replicate the underlying structural topology, nor do we expect them to \textit{in general} (particularly in neuroimaging experiments, where aspects of the structure may be expressed more or less or not at all, depending on the cognitive task).
In spite of that, in this paper we do seek to evaluate and indeed validate these methods in inferring microscopic, mesoscopic and macroscopic features of the underlying network structure.
Crucially, we perform this validation under idealised conditions -- including full observability, stationarity no subsampling, etc. -- which allow us to establish a hypothesis that effective networks should be not just complementary to the structural but converge to it under these conditions, as our available data increases.
Indeed, under these idealised conditions (specifically in the absence of hidden nodes, and other simplifying assumptions, including stationarity), effective networks inferred via multivariate TE are proven to converge to the underlying structure for sufficiently long time series~\citep{Sun2015,Runge2018chaos}.
In gaining an understanding of these multivariate effective connectivity inference algorithms, it is important to validate that they perform to that expectation where it is applicable, and investigate how that performance varies with respect to factors such as sample size, etc.
In doing so, we also address the recent call for more extensive model diversity in testing multivariate algorithms: ``\textit{To avoid biased conclusions, a large number of different randomly selected connectivity structures should be tested [including link density as well as properties such as small-worldness]}''~\citep{Runge2018chaos}.

Outside of these idealised circumstances though, we can no longer make a clear general hypothesis on how the effective network models are expected to reflect the underlying structure, yet a successful validation gives confidence that the directed statistical relationships they represent remain accurate as an effective network model at the microscale.
Furthermore, it is at least desirable for not only effective networks but also functional networks to recognise important features in the underlying network structure: to track overall regime changes in the macroscopic structure reliably, and to reflect the mesoscopic properties of distinctive nodes (or groups of nodes) in the structure.
The desire for recognition of important features in the network is applicable whether the inference is made under idealised conditions or not.

This motivates our validation study, which is primarily conducted under idealised conditions as above and based on synthetic datasets involving ground truth networks of \num{100}--\num{200} nodes with different topologies, from regular lattice to small-world\ifarXiv~~and\fi random\ifarXiv (\secRef{WS})\else\fi, scale-free\ifarXiv (\secRef{BA})\else\fi, and modular\ifarXiv (\secRef{modular})\else\fi.
Many of these structural properties are incorporated in the macaque connectomes analysed in \ifarXiv\secRef{macaque}\fi.
At the macroscale, we measure several fundamental and widely-used properties, including shortest-path length, clustering coefficient, small-world coefficient, betweenness centrality, and features of the degree distributions~\citep{Rubinov2010}.
These properties of the inferred network models are compared to those of the real underlying structural networks in order to validate and benchmark different inference algorithms in terms of their ability to capture the key properties of the underlying topologies.
At the microscale, the performance is assessed in terms of precision, recall, and specificity of the inferred model in classifying the links of the underlying structural network.
As above, whilst we do not expect all approaches to strictly capture the microscale features, these results help to explain their performance at the macroscale.

For most of our experiments, the time series of node activity on these networks are generated by vector autoregressive (VAR) dynamics, with linearly coupled nodes and Gaussian noise.
Both the VAR process and the inference algorithms are described in detail in \ifarXiv\secRef{methods}\fi, where we also discuss how MI and the magnitude of Pearson correlation are equivalent for stationary VAR processes.
This implies that the undirected networks obtained via the bivariate MI algorithm are equivalent to the widely employed undirected functional networks obtained via correlation, extending the implications of our results beyond information-theoretic methods.
Further, our results based on TE extend to Granger causality, which is equivalent to TE for stationary VAR processes~\citep{Barnett2009}.
Networks inferred using bivariate TE are typically referred to as directed functional networks, to emphasise the directed nature of their links.
The extension to multivariate TE for effective network inference can also be viewed as an extension to multivariate Granger causality for the stationary VAR processes here.

We find that multivariate TE performs better on all network topologies at all scales, for longer time series.
Bivariate methods can achieve better recall with limited amount of data (shorter time series) in some circumstances, but the precision and the ability to control false positives are not consistent nor predictable a priori.
On the other hand, thanks to recent statistical improvements, multivariate TE guarantees high specificity and precision regardless of the amount of data available, and the recall steadily increases with more data.
We discuss how the mesoscopic properties of the underlying structural network---particularly the network motifs---can influence the precision and recall of the model at the microscale.
In turn, we show how the performance at the microscale affects the inferred network properties at the macroscale. We observe that bivariate methods are often unable to capture the most distinctive topological features of the networks under study (including path length, clustering, degree distribution, and modularity), largely due to their inability to control false positives at the microscale.

Our final section moves beyond the validation under idealised conditions to extend the experiments to time series from a neural mass model employed on the CoCoMac connectome.
Although the incorporation of complexities such as nonlinear interactions and subsampled time series do somewhat reduce the performance of the methods, the superior performance of multivariate TE aligns with the results above from the validation in idealised conditions.
Whilst further and more wide ranging experiments are required, our experiments provide substantial evidence for the validity of multivariate TE in providing effective network \textit{models}, which still retain meaningful network insights in more realistic conditions.

\section{Methods}
\label{sec:methods}

\subsection{Generating dynamics on networks}
\begin{figure*}
    \ifarXiv\includegraphics[width=\textwidth]{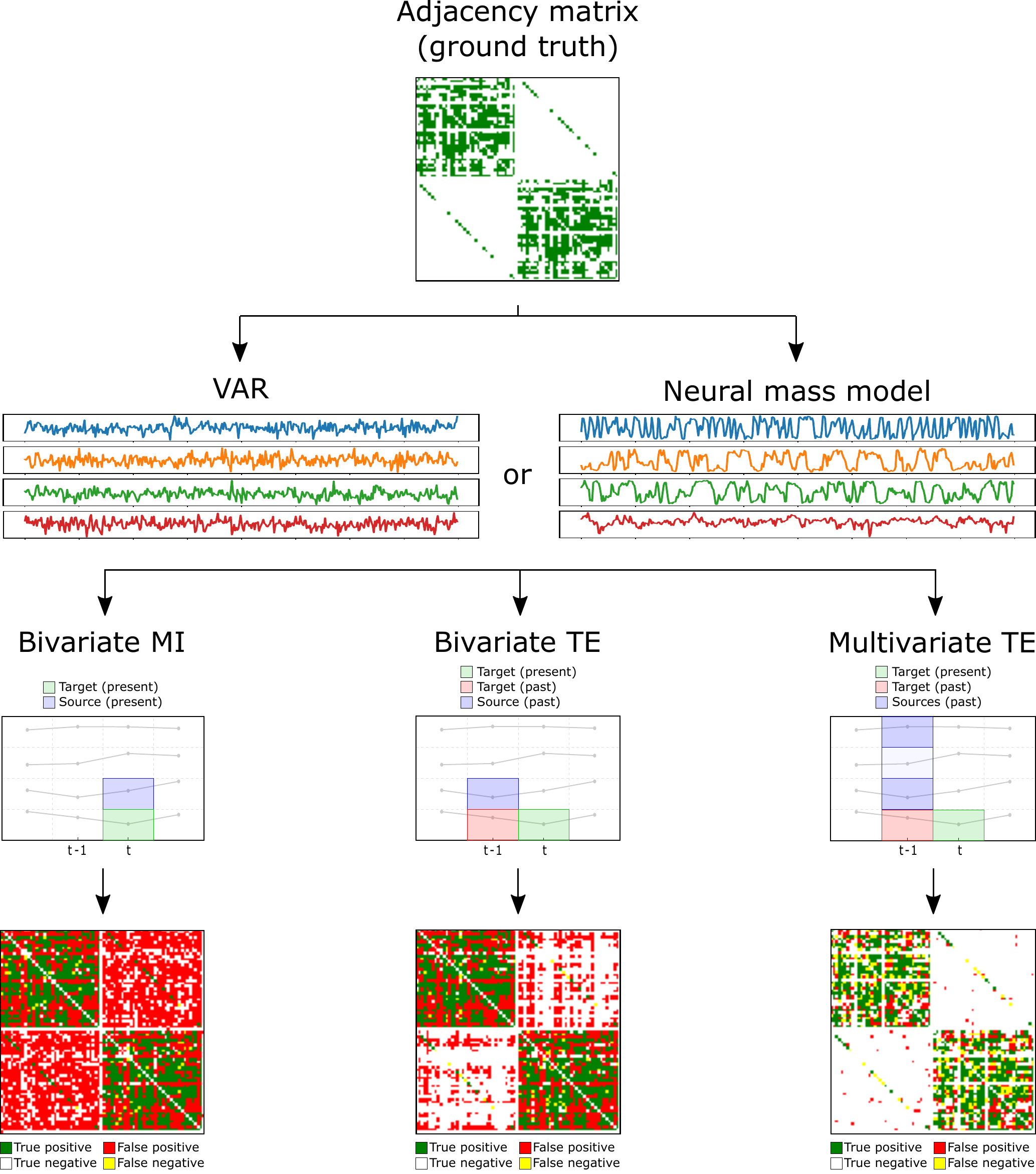}
    \fi
    \caption{\label{fig:pipeline}
        Pipeline of the network inference comparison, using the CoCoMac connectome as an example (see \textit{Methods}~\ifarXiv\secRef{methods}\fi and \textit{Numerical simulations} in \ifarXiv\secRef{macaque}\fi~for full details).
        With the real adjacency matrix defining the links, synthetic time series are generated using either a linear autoregressive (VAR) system with uniform link weights, or a nonlinear neural mass model with realistic weights.
        Three network inference algorithms (bivariate MI, bivariate TE, multivariate TE) are then employed to analyse the time series.
        Bivariate methods consider pairs of nodes independently of each other, either taking the past states into account (bivariate TE) or not (bivariate MI).
        On the other hand, multivariate TE analyses pairs of nodes in the context of the whole network.
        At the microscale, the links in the inferred networks are classified as true/false positives/negatives and these scores are used to compute standard performance measures (precision, recall, and specificity).
        At the macroscale, the performance is instead measures according to the ability of each algorithm to faithfully reflect network properties (summary statistics) of the underlying structural network serving as ground truth.
    }
\end{figure*}

Two models are used to generate time series dynamics on networks of coupled variables. 
Vector autoregressive processes are employed for validation studies under idealised conditions, and a neural mass model is used on the weighted CoCoMac connectome as a final investigation going beyond these conditions.
The simulation and analysis pipeline is illustrated in \fig{pipeline}.

\subsubsection{Networks of linearly-coupled Gaussian variables}
\label{sec:VAR}
We simulate discrete-time, stationary, first-order vector autoregressive processes (VAR) on underlying structural networks of $N$ nodes.
A VAR process is described by the recurrence relation
\begin{equation}\label{eq:VAR}
    \bm{Z}(t+1)=\bm{Z}(t)\cdot C+\bm{\varepsilon}(t),
\end{equation}
where $\bm{Z(t)}$ is a row vector and $Z_i(t)$ is the activity of node~$i$ at time~$t$.
The Gaussian noise $\bm{\varepsilon}(t)$ is spatially and serially uncorrelated, with standard deviation $\theta=0.1$.
The $N\times N$ weighted adjacency matrix $C=~[C_{ij}]$ describes the network structure, where $C_{ij}$ is the weight of the directed connection from node~$i$ to node~$j$.
These dynamics are generated on various network topologies, as detailed in the following sections.
The choice of the weights $C_{ij}$ (detailed in the following sections) guarantees the stability of the system, which is a sufficient condition for stationarity~\citep{Atay2006}.
Since stationary VAR processes have multivariate Gaussian distributions, the information-theoretic measures we use can be directly related to Pearson correlation and Granger causality~\citep{Granger1969}.
The simple VAR dynamics is chosen as the primary model for our validation studies instead of nonlinear alternatives because the main goal is not to prove the superiority of nonlinear dependence measures on nonlinear systems, which has been shown elsewhere~\citep{Novelli2019}.
We rather aim to show that, even on linearly-coupled Gaussian variables---perfectly suitable to be studied via cross-correlations---multivariate approaches are better able to infer the macroscopic network properties.
In addition, the VAR dynamics is amenable to be investigated using the faster Gaussian estimator for the information-theoretic measures, allowing us to carry out more extensive simulations over a wider range of parameters.

\subsubsection{Neural mass model on the CoCoMac connectome}
To provide an extension beyond the linear VAR dynamics, neural activity in various brains regions is modelled (following~\citet{Shine2018} and~\citet{Li2019}) as an oscillating 2-dimensional neural mass model derived by mode decomposition from the Fitzhugh-Nagumo single neuron model \cite{FitzHugh1961}.
As previously presented \citep{Shine2018,Li2019}, the CoCoMac connectome~\citep{Kotter2004} is used to provide directed coupling between \num{76} regions, with axonal time delays between these regions based on the length of fibre tracts as estimated by diffusion spectrum imaging \citep{SanzLeon2013}.
Data provided from \citep{Shine2018,Li2019} was simulated using the open source framework The Virtual Brain~\citep{SanzLeon2013}, using code implementing the model freely available at \href{https://github.com/macshine/gain_topology}{https://github.com/macshine/gain\_topology} \cite{ShineCode2018}.

Langevin  equations \eq{neural_mass_model} specify the neural mass model, via the dynamics of local mean membrane potential ($V$) and the slow recovery variable ($W$) at each regional node $i$:
\begin{align}
    \label{eq:neural_mass_model}
    \dot{V}_i(t) &= 20 \big( W_i(t) + 3V_i(t)^2 - V_i(t)^3 + \gamma I_i \big) + \xi_i(t), \nn \\ 
    \dot{W}_i(t) &= 20 (- W_i(t) - 10V_i(t)) + \eta_i(t).
\end{align}
In the above, $\xi_i$ and $\eta_i$ are independent standard Wiener noises and $I_i$ is the synaptic current
\begin{align}
I_i = \sum_{j} C_{ji} S_j (t - \tau_{ji}),
\end{align}
with $C_{ji}$ indicating the connection weight from $j$ to $i$ and incorporating time delays $\tau_{ji}$ from $j$ to $i$ (estimated as described above).
The CoCoMac connectome network contains $1560$ directed connections (including $66$ self-links), with $\tau_{ji}$ on non-self links having an average of $19.8$ ms (standard deviation $8.32$ ms).
The membrane potentials $V_i$ are converted to normalised firing rates $S_i$ via a sigmoid activation function
\begin{align}
	S_i(t) = \frac{1}{1 + \textit{e}^{-\sigma(V_i(t) - m)}},
	\label{eq:S}
\end{align}
with parameter $m = 1.5$ chosen to align the sigmoid with its typical input.
The parameters for gain $\sigma=0.5$ (in \eq{S}) and excitability $\gamma=0.3$ (in \eq{neural_mass_model}) are selected to simulate activity in the integrated regime of dynamics identified by~\citet{Shine2018}.

Finally, the time series of membrane voltage $V_i(t)$ (originally obtained with a $0.5$ ms temporal resolution via stochastic Heun integration) are subsampled at $15$ ms (selected as half the median time for the autocorrelation functions to decay to $1/e$).

\subsection{Network inference algorithms}
As illustrated in \fig{pipeline}, three algorithms are employed to infer network models from the time series using the IDTxl Python package~\citep{Wollstadt2019}:

\subsubsection{Bivariate mutual information for functional connectivity\label{sec:methods_bMI}}
Mutual information (MI) is computed between all pairs of nodes independently, in a \textit{bivariate} fashion, and only the measurements that pass a strict statistical significance test (described below) are interpreted as undirected links.

MI is a measure of statistical dependence between random variables~\citep{Cover2005}, introduced by Shannon in laying the foundations of information theory~\citep{Shannon1948}.
Formally, the MI between two continuous random variables $X$ and $Y$ with joint probability density function $\mu(x,y)$ and marginal densities $\mu_X(x)$ and $\mu_Y(y)$ is defined as
\begin{equation}
    I(X;Y) \coloneqq \iint \mu(x,y) \log \frac{\mu(x,y)}{\mu_X(x)\mu_Y(y)} dx dy,
\end{equation}
where the integral is taken over the set of pairs $(x,y)$ such that $\mu(x,y)>0$.
The strength of MI lies in its model-free nature, meaning that it doesn't require any assumptions on the distribution or the variables (\eg Gaussian).
Being able to capture nonlinear relationships, MI is typically presented as a generalised version of the Pearson correlation coefficient.
However, for the VAR processes considered here [\eq{VAR}] with stationary multivariate Gaussian distributions, the MI between two variables $X$ and $Y$ is completely determined by the magnitude of their Pearson correlation coefficient $\rho$~\citep{Cover2005}:
\begin{equation}
    I(X;Y)=-\ln(1-\rho^2).
\end{equation}
Crucially, this one-to-one relationship between MI and the absolute value of $\rho$ (for VAR processes) implies that the networks inferred via the bivariate MI algorithm are equivalent to the functional networks obtained via cross-correlation---widely employed in neuroscience.
This equivalence persists whenever a Gaussian estimator for MI is used (which models the processes as VAR), even for nonlinear dynamics, as is used in our experiments.
Differences may lie in how the raw MI values are transformed into a network structure.
Early approaches often used a fixed threshold aimed at obtaining a prescribed link density, while the bivariate MI algorithm used here adopts an adaptive threshold (different for each link) to meet a desired statistical significance level.
The statistical significance is computed via null hypothesis testing to reflect the probability of observing a larger MI from the same samples if their temporal relationship were destroyed (the $p$-value is obtained from a chi-square test, as summarised in~\citep{Lizier2014JIDT}).
The critical level for statistical significance is set to $\alpha=0.01/N$, where $N$ is the network size.
This produces a Bonferroni correction for the inference of parent nodes for each target (\ie for each target, there is a $0.01$ chance under the null hypothesis that at least one spurious parent node is selected, assuming independent sources).

\subsubsection{Bivariate transfer entropy for directed functional connectivity}
Transfer entropy (TE) is computed between all pairs of nodes independently, in a \textit{bivariate} fashion, and only the measurements that pass a strict statistical significance test (described below) are interpreted as links.

TE is a model-free measure of statistical dependence between random variables~\citep{Schreiber2000}; however, differently from MI and cross-correlation, it is a directed and not symmetric measure (\ie the TE from a source node $X$ to a target node $Y$ is not necessarily the same as the TE from $Y$ to $X$), and specifically considers information about the dynamic state updates of the target $Y$.
Thus, employing TE has the advantage of generating directed networks and providing a more detailed model of the dynamics of the system under investigation.
Formally, the TE from a source stochastic process $X$ to a target process $Y$ is defined as~\citep{Schreiber2000}
\begin{equation}
    T_{X\rightarrow Y}(t) \coloneqq I(X_{t-1};Y_{t}|\bm{Y}_{<t}),
\end{equation}
where $I(X_{t-1};Y_{t}|\bm{Y}_{<t})$ is the conditional mutual information~\citep{Cover2005} between the previous sample $X_{t-1}$ of the source and the next sample $Y_{t}$ of the target, in the context of (the vector of) the target's past values $\bm{Y}_{<t}$.
The directed and dynamic nature of TE derives specifically from taking the past of the target into account when measuring the lagged statistical dependence between $X$ and $Y$. 

In practice, in order to estimate the TE from the time series, $\bm{Y}_{<t}$ is usually constructed as an embedding vector~\citep{Takens1981} and a maximum lag must be specified to build a finite embedding of the target's past (either using uniformly- or non-uniformly-spaced variables~\citep{Vlachos2010,Faes2011,Kugiumtzis2013}).
Here, using the IDTxl Python package~\citep{Wollstadt2019}, a non-uniform embedding of the target's past $\bm{Y}_{<t}$ is built via iterative greedy selection of statistically significant elements of $\bm{Y}_{<t}$ (as detailed in the next section).
Whilst similar multivariate embeddings and lags larger than one time step can be used for the source $X$ in principle, here only a single sample at lag $1$ for the source is used for the VAR model in alignment with its dynamics.

Analogously to the MI, the bivariate TE values are transformed into a directed network structure by testing their statistical significance.
This is computed (using a theoretical null-distribution, as summarised in~\citep{Lizier2014JIDT}) to reflect the probability of observing a larger TE from the same samples if the source samples were temporally decoupled from the target and its past.
A critical level of $\alpha=0.01/N$ is used as per MI.

Bivariate TE has found wide application in studies of directed functional connectivity, \eg~\citep{Honey2007,Stetter2012,Orlandi2014,Marinazzo2012,Lizier2011,Wibral2011,Wibral2014}.
Importantly, TE and Granger causality are equivalent for Gaussian variables~\citep{Barnett2009}, which applies to the VAR processes considered here [\eq{VAR}], suggesting a
Gaussian estimator for TE be employed~\citep{Bossomaier2016}. The networks inferred via the bivariate TE algorithm with this estimator are equivalent (for any dynamics) to those obtained via Granger causality, which is also widely employed in neuroscience.

\subsubsection{Multivariate transfer entropy for effective connectivity}
Differently from the bivariate approaches above, the multivariate TE approach does not consider pairs of nodes in isolation, but focusses on modelling dynamic updates in each target process in the system by selecting a \emph{minimal set} of sources that collectively contribute to the computation of the target's next state.
More formally, for each target $Y$, this method aims at identifying the minimal set of sources $\bm{X}_{<t}$ that maximise the collective TE to $Y$, defined as
\begin{equation}
    T_{\bm{X}\rightarrow Y}(t) \coloneqq I(\bm{X}_{<t};Y_{t}|\bm{Y}_{<t}).
\end{equation}
The multivariate TE network inference algorithm is described in full in~\citep{Novelli2019}, synthesising together components of~\citep{Vlachos2010,Faes2011,Lizier2012,Montalto2014,Sun2015}.
A greedy approach is used to iteratively select the candidate variables to add to $\bm{X}_{<t}$ from a candidate set $\{X_{i,t-l}\}$ of lagged variables from the past of each source $X_i \in \mathbf{Z} \setminus \{Y\}$, up to some maximum lag $l \leq L$ ($L=1$ for the VAR model and $L=4$ for the neural mass model used here).
Testing the statistical significance of the conditional mutual information for a new candidate source to be included at each step---conditioning on the previously selected sources---provides an adaptive stopping condition for the greedy algorithm.
The family-wise error rate for each target is set to $\alpha=0.01$ using the max statistic~\citep{Novelli2019}, meaning that for each target there is a $0.01$ chance that, under the null hypothesis, at least one spurious parent node is selected.
The conditioning on previously selected sources serves to prevent the selection of spurious sources which are correlated with true sources due to common driver as well as pathway or chain effects (referred to as holding redundant information only~\citep{Lizier2012,Stramaglia2014}). Such conditioning also enables capturing multivariate or synergistic effects on the target that cannot be detected by examining individual sources in isolation~\citep{Lizier2012,Stramaglia2014}.
Furthermore, in contrast to always conditioning on all potential sources, by conditioning \textit{only} on previously selected sources the iterative approach defers overly-high dimensional analysis until it is genuinely required, buying statistical sensitivity (``recall'')~\citep{Novelli2019,Lizier2012}. Every node is studied as a target (in parallel on a computing cluster, using IDTxl~\citep{Wollstadt2019}) and the results are then combined into a directed network describing the information flows in the system.
Similarly to the bivariate TE discussed above, a non-uniform embedding of the target's past $\bm{Y}_{<t}$ is built first~\citep{Vlachos2010}, before the second step of selecting sources via the same iterative greedy algorithm~\citep{Novelli2019}.
Whilst multiple past samples of any given source can been considered (\eg as has been done in~\citep{Novelli2019}), only one past value is examined here ($L=1$) for the VAR experiments in line with their known structure in \eq{VAR} and in order to focus on network structure effects only.

Given that TE and Granger causality are equivalent for Gaussian variables~\citep{Barnett2009}, using the Gaussian estimator with the multivariate TE algorithm can be viewed as extending Granger causality in the same multivariate/greedy fashion.

\subsection{Evaluation metrics}
At the microscale (individual links), the network inference performance is evaluated against the known underlying network structure as a binary classification task, using standard statistics based on the number of \emph{true positives}~(TP, \ie correctly classified existing links), \emph{false positives}~(FP, \ie absent links falsely classified as existing), \emph{true negatives}~(TN, \ie correctly classified absent links), and \emph{false negatives}~(FN, \ie existing links falsely classified as absent).
The following standard statistics are employed in the evaluation:
\begin{description}[]
\item[Precision]
$=TP/(TP+FP)$
\item[Recall (true-positive rate)]
$=TP/(TP+FN)$
\item[Specificity (true negative rate)]
$=TN/(TN+FP)$
\item[False-positive rate]
$FP/(FP+TN)$
\end{description}
Intuitively, the precision measures how often an inferred link is actually present in the underlying structure, the recall measures the proportion of true links that are detected, and the specificity measures the proportion of absent links that are correctly not inferred.
For a properly controlled family-wise error rate ($\alpha$), the expected specificity is $1-\alpha$.

At the macroscale, the performance is evaluated in terms of the accuracy in measuring network properties of interest on the inferred network, as compared to their real values when measured on the underlying structural network.
These properties include:
\begin{description}[]
\item[Characteristic path length]
The average shortest distance between all pairs of nodes~\citep{Rubinov2010}.
A shorter average value is typically interpreted as an indication of the efficiency of the network in propagating information.
The characteristic path length is only well defined for connected networks---a problem that is avoided by construction in this study by only generating connected networks.
This limitation could be alternatively overcome by replacing the characteristic path length with the analogous global efficiency~\citep{Latora2001}, reported in \ifarXiv\appRef{efficiency}\fi for completeness.
\item[Clustering coefficient]
The clustering coefficient of a node is the fraction of the node's neighbours that are also neighbours of each other~\citep{Watts1998,Fagiolo2007}.
The mean clustering coefficient hence reflects the average prevalence of clustered connectivity around individual nodes.
\item[Small-worldness coefficient]
Small-world networks are formally defined as networks that are significantly more clustered than random networks, yet have approximately the same characteristic path length as random networks~\citep{Watts1998}.
The small-worldness coefficient was proposed by \citet{Humphries2008} to capture this effect in the following single statistic (although improvements on the original measure have been recently suggested~\citep{Neal2017,Zanin2015}):
\begin{equation}
\label{eq:small_worldness_coefficient}
    \sigma=\frac{\frac{C}{C_{\textup{rand}}}}{\frac{L}{L_{\textup{rand}}}}.
\end{equation}
In \eq{small_worldness_coefficient}, $C$ and $L$ respectively denote the average clustering coefficient and the characteristic path length; analogously, $C_{\textup{rand}}$ and $L_{\textup{rand}}$ denote the average clustering coefficient and the characteristic path length of Erd\H{o}s-R\'{e}nyi networks having the same size and number of links as the network under study.
\item[Degree distribution]
The probability distribution of the in- and out-degree over the whole network (where the in- and out-degree of a node are defined as the number of its incoming and outgoing connections, respectively).
\item[Betweenness centrality]
The fraction of all shortest paths in the network that pass through a given node~\citep{Rubinov2010}, excluding paths that start and end on the given node.
\item[Modularity]
A measure of the separation of a network into specific groups (or modules), defined as the fraction of the edges that fall within the given groups minus the expected fraction if the edges were distributed at random~\citep{Newman2004}.
\item[Rich-club coefficient]
The ``rich-club'' phenomenon refers to the tendency of hubs (nodes with high degree) to form tightly interconnected communities~\citep{Colizza2006}.
The rich-club coefficient was first proposed by \citet{Zhou2004} to quantify this effect:
\begin{equation}
\label{eq:rich_club_coefficient}
    \phi(k)=\frac{2 E_{>k}}{N_{>k}(N_{>k}-1)},
\end{equation}
whereby $E_{>k}$ denoted the number of edges among the $N_{>k}$ nodes having
degree higher than a given value $k$.
\end{description}

\section{Small-world networks}
\label{sec:WS}

\subsection{Numerical simulations}
The first experiment is aimed at testing the robustness of the three inference algorithms with respect to vast changes in network structure.
The Watts-Strogatz model is used to generate a spectrum of topologies, ranging from regular lattices to random networks (similar to Erd\H{o}s-R\'{e}nyi networks, although not equivalent~\citep{Maier2019}) through a small-world transition~\citep{Watts1998}.
Each simulation starts with a directed ring network of \num{100} nodes with uniform link weights $C_{ij}=C_{ii}=0.15$ and fixed in-degree~$d_\textnormal{in}=4$ (\ie each node is linked to two neighbours on each side, as well as to itself via a self-loop).
The source of each link is then rewired with a given probability $p \in [0,1]$, so as to change the overall network topology while keeping the in-degree of each node fixed.
Only rewiring attempts that keep the network connected are accepted, in order to allow the measurement of the average shortest-path length.
The simulations for each $p$ are repeated \num{10} times on different network realisations and with random initial conditions.

\subsection{Results}
\begin{figure}
    \ifarXiv\includegraphics[width=0.48\textwidth]{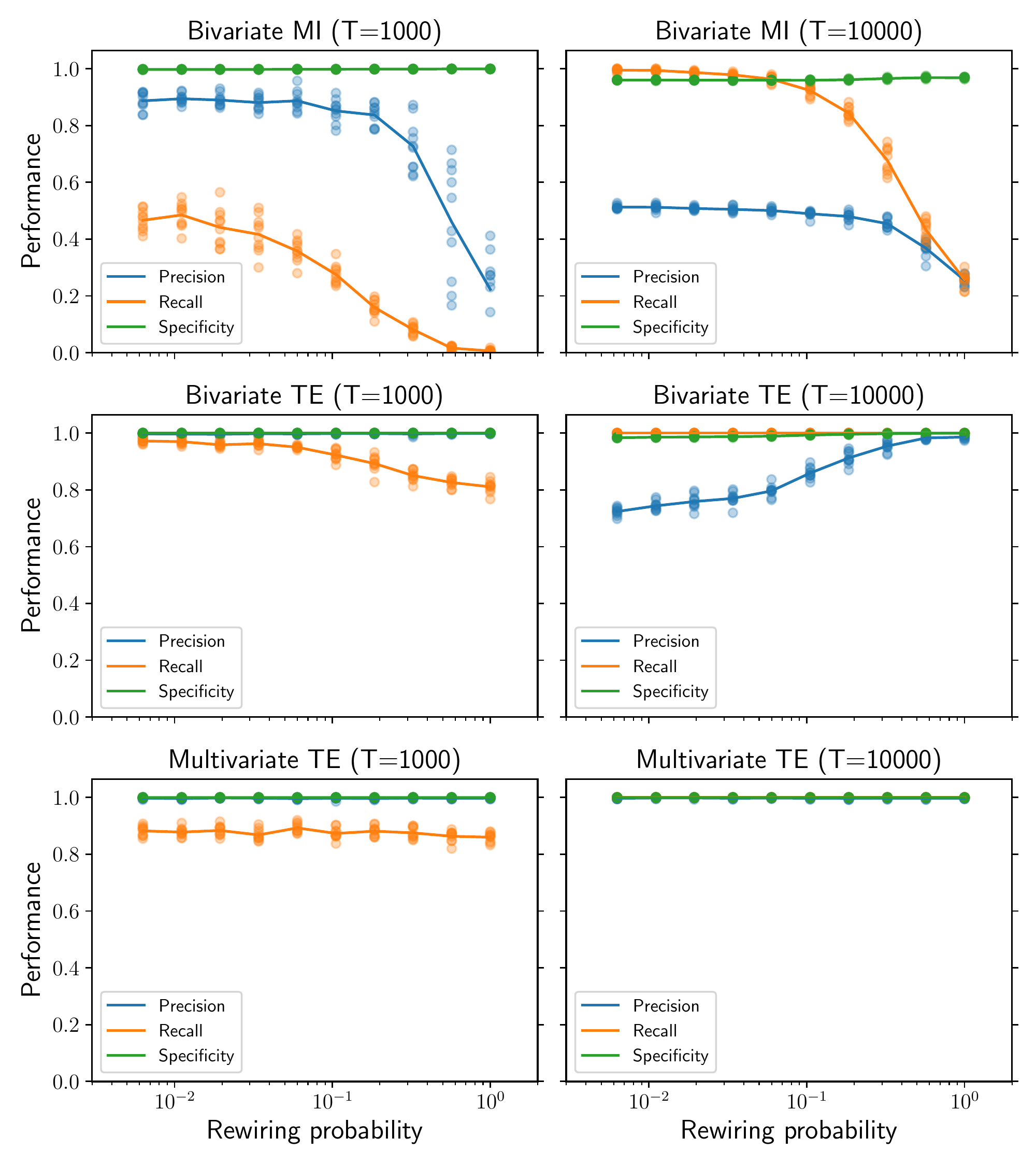}
    \fi
    \caption{\label{fig:WS_performance_vs_rewiring}
        Performance as a function of the rewiring probability in Watts-Strogatz ring networks ($N$=\num{100} nodes).
        Multivariate TE is consistent across network structure and guarantees high precision regardless of the amount of data available (third row).
        Bivariate TE (second row) can have better recall than multivariate TE for shorter time series ($T=1000$, left column) but its precision is not consistent (it drops when $T$=\num{10000}, right column) and the optimal time series length cannot be determined a priori.
        Bivariate MI (first row) has lower precision and recall than TE-based methods, both for $T$=\num{1000} and $T$=\num{10000}.
        For each value of the rewiring probability, the results for~\num{10} simulations on different networks are presented (low-opacity markers) in addition to the mean values (solid markers).
    }
\end{figure}

At the microscale, the performance is evaluated in terms of precision, recall, and specificity in the classification of the links (present or absent) in the inferred network compared to the underlying structural network.
In the case of bivariate MI, each undirected link in the inferred network is represented as two directed links in opposite directions.
For longer time series of \num{10000} samples, multivariate TE is the most accurate method, achieving optimal performance according to all metrics on all the network topologies generated by the Watts-Strogatz rewiring model (\fig{WS_performance_vs_rewiring}, right column).
Bivariate TE also achieves nearly optimal recall and high specificity on all topologies; however, despite the strict statistical significance level, the precision is significantly lower on lattice-like topologies (low rewiring probability) than on random ones (high rewiring probability).
The opposite trend is shown by the bivariate MI algorithm, whose precision and recall drastically decrease with increasing rewiring probability.
As expected, the recall of all methods decreases when shorter time series of \num{1000} samples are provided (\fig{WS_performance_vs_rewiring}, left column).
However, the recall for multivariate TE is consistent across topologies, while it decreases with higher rewiring probability when bivariate methods are used.
This results in the bivariate TE having larger recall for lattice-like topologies whilst multivariate TE has larger recall than bivariate for more random topologies (\ie for a rewiring probability larger than $p=0.2$).
A further interesting effect is that bivariate TE attains better precision on shorter time series than on longer ones.

\begin{figure}
    \ifarXiv\includegraphics[width=0.48\textwidth]{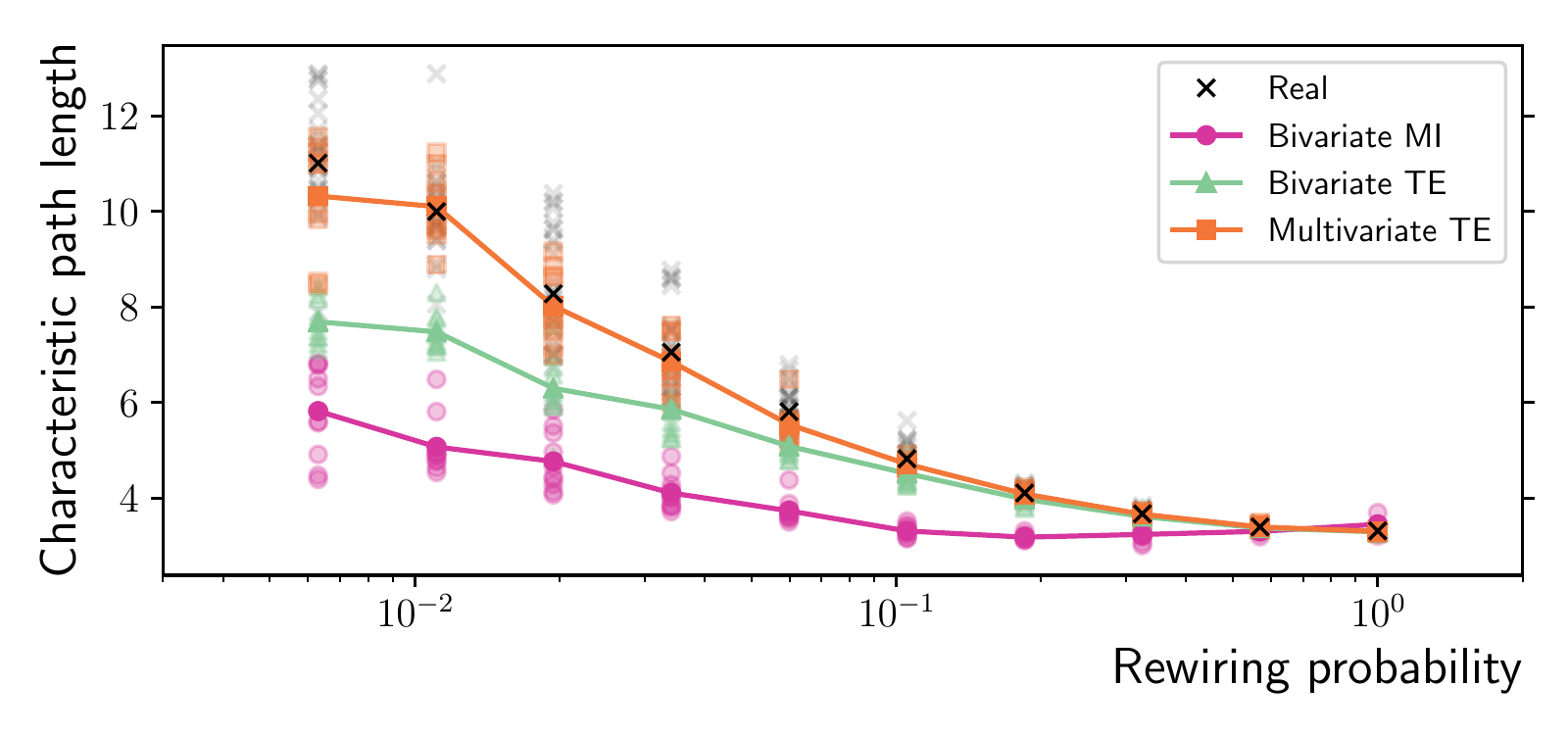}
    \fi
    \caption{\label{fig:WS_shortest_path_vs_rewiring_T10000}
        Characteristic path length as a function of the rewiring probability in Watts-Strogatz ring networks ($N$=\num{100} nodes and $T$=\num{10000} time samples).
        Multivariate TE is able to closely approximate the characteristic path length of the real topologies (ground truth).
        On the other hand, bivariate MI and TE produce significant underestimates due to spurious links creating shortcuts across the network, particularly on lattice-like topologies (low rewiring probability).
        The results for~\num{10} simulations on different network realisations are presented (low-opacity markers) in addition to the mean values (solid markers).
    }
\end{figure}

\begin{figure}
    \ifarXiv\includegraphics[width=0.48\textwidth]{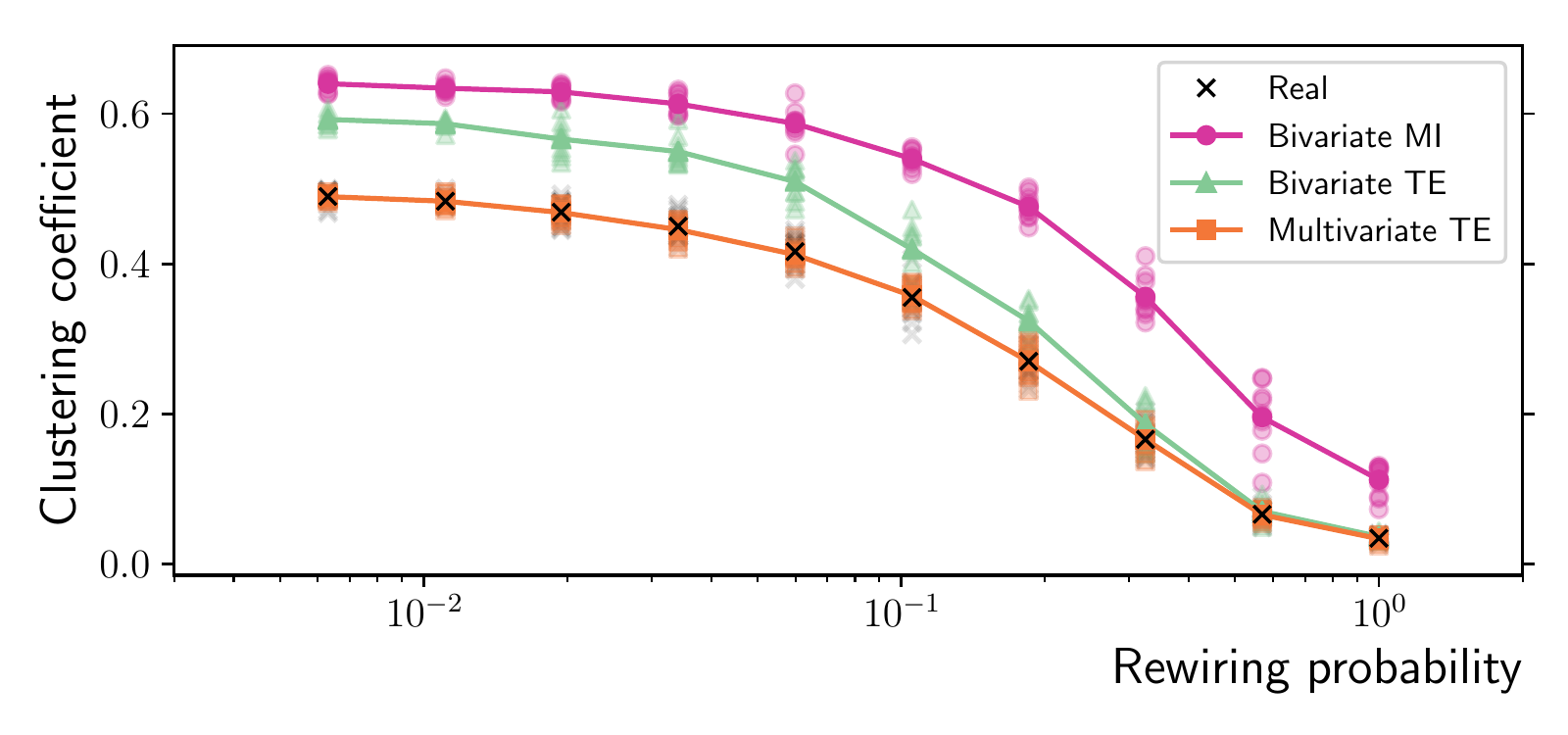}
    \fi
    \caption{\label{fig:WS_clustering_vs_rewiring_T10000}
        Average clustering coefficient as a function of the rewiring probability in Watts-Strogatz ring networks ($N$=\num{100} nodes and $T$=\num{10000} time samples).
        The multivariate TE algorithm closely matches the average clustering coefficient of the real networks (ground truth), which is instead overestimated by bivariate MI and TE.
        The results for~\num{10} simulations on different network realisations are presented (low-opacity markers) in addition to the mean values (solid markers).
    }
\end{figure}

At the macroscale, the three algorithms are tested on their ability to accurately measure three fundamental network properties relevant through the small-world transition, using the longer time series of \num{10000} samples.
Multivariate TE is able to closely approximate the real shortest-path length on all the network topologies generated by the Watts-Strogatz rewiring model, while the bivariate MI and TE algorithms produce significant underestimates, particularly on lattice-like topologies (\fig{WS_shortest_path_vs_rewiring_T10000}).
Similarly, multivariate TE is able to closely match the real mean clustering on all the network topologies generated by the Watts-Strogatz rewiring model, while the bivariate MI and TE algorithms consistently overestimate it (\fig{WS_clustering_vs_rewiring_T10000}).
The related measure of local efficiency~\citep{Latora2001} is reported in \ifarXiv\appRef{efficiency}\fi.

\begin{figure}
    \ifarXiv\includegraphics[width=0.48\textwidth]{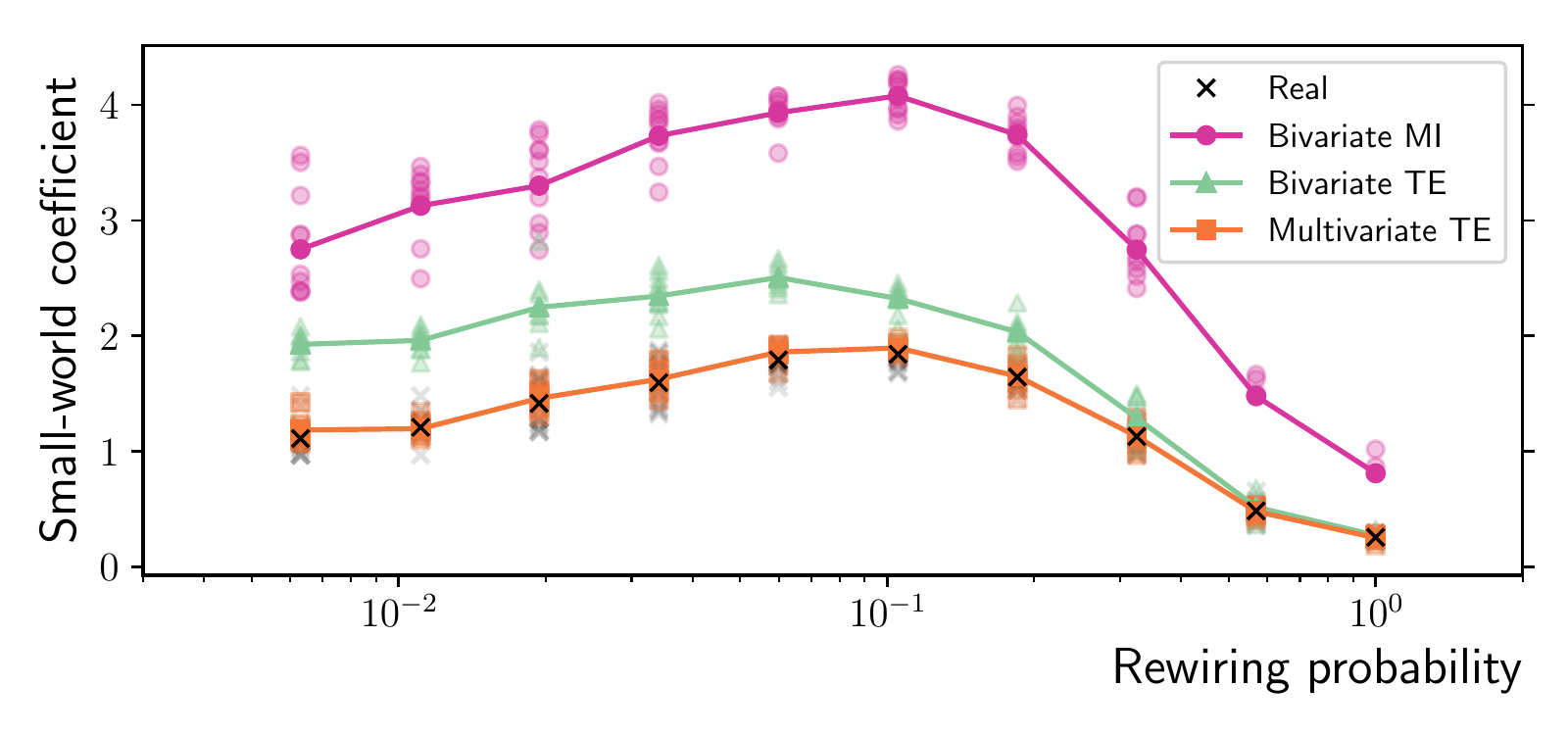}
    \fi
    \caption{\label{fig:WS_SW_coeff_vs_rewiring_T10000}
        Small-world coefficient as a function of the rewiring probability in Watts-Strogatz ring networks ($N$=\num{100} nodes and $T$=\num{10000} time samples).
        The multivariate TE algorithm produces accurate estimates of the small-world coefficient of the real topologies (ground truth), which is instead strongly overestimated by bivariate MI and TE.
        The results for~\num{10} simulations on different network realisations are presented (low-opacity markers) in addition to the mean values (solid markers).
    }
\end{figure}

Given the above results on the characteristic path length and the mean clustering coefficient, it is not surprising that the bivariate MI and TE algorithms significantly overestimate the real small-worldness coefficient, while the multivariate TE method produces accurate estimates on all the network topologies generated by the Watts-Strogatz rewiring model (\fig{WS_SW_coeff_vs_rewiring_T10000}).
Equivalent results are found if the alternative measures of ``small-world index"~\citep{Neal2017} or ``double-graph normalized index"~\citep{Telesford2011} are computed instead (not shown).

\subsection{Discussion}
At the microscale, the results concerning the bivariate TE can be explained in the light of the recent theoretical derivation of TE from network motifs for VAR dynamics~\citep{Novelli2020}.
For a fixed in-degree, the TE decreases with the rewiring probability, making it harder for candidate links to pass the statistical significance tests when only short time series are available.
This explains why the recall for the bivariate TE slightly drops with higher rewiring probability for $T$=\num{1000} (\fig{WS_performance_vs_rewiring}).
We can speculate that a similar mechanism could be responsible for the more drastic drop in the recall for the bivariate MI (via evidence from derivations for covariances from network structure for similar processes~\citep{Pernice2011,Schwarze2020}).
The fact that bivariate TE is larger for regular lattice structures also explains why its recall is slightly higher than for multivariate TE here: the redundancy between close sources that elevates their bivariate TE is explicitly conditioned out of the multivariate TE for secondary sources.
On the other hand, as the rewiring increases, the higher recall for multivariate TE must be due to this method capturing synergistic effects that (more disparate) multiple sources have on the target, which the bivariate method does not.

Comparing the results between shorter and longer time series raises another question: why is the precision of the bivariate TE worse for longer time series than for shorter ones, especially for lattice-like topologies?
More complex motifs involving common parents and multiple walks, which are more prevalent in regular lattice topologies, can result in nonzero TE on spurious links.
These indirect effects are typically weak; however, for long enough time series, the low TE values can be distinguished from noise and thus pass the statistical significance tests.
The resulting spurious links (false positives) decrease the precision and the specificity as the time series length is increased, with the effect being stronger in regular lattice topologies.
In other words, the Bonferroni correction of the statistical significance level (\ie dividing $\alpha$ by the network size $N$) does not result in a well calibrated test for bivariate inference methods---the sources are correlated, and the tests on them are not independent.
The differences in the specificity on the plots are subtle because the networks are sparse; however, they manifest in large differences in the precision.
Crucially, this effect is not seen for the multivariate TE, which maintains specificity consistent with the requested $\alpha$ for all topologies and time series lengths.
Thus, lower recall achieved by multivariate TE on regular lattice networks for short time series (compared to bivariate TE) can be viewed as a compromise to control the specificity in a consistent fashion.
A compelling argument in favour of controlling the specificity is provided by~\citet{Zalesky2016}, who conclude that ``\textit{specificity is at least twice as important as sensitivity [\ie~recall] when estimating key properties of brain networks, including topological measures of network clustering, network efficiency and network modularity}''.
Unfortunately, there is currently no consistent \textit{a priori} way (nor a reasonable candidate) to determine the optimal time series length for bivariate TE to attain high precision.

Moving to the macroscale results, it is clear that the ability to control the false positives while building connectomes is a crucial prerequisite for the application of complex network measures.
Adding only a few spurious links leads to significant underestimate of the average shortest-path length---an effect that has previously been reported for lattice-like networks using MI~\citep{Bialonski2010} and extended here to TE and across a range of topologies (\fig{WS_shortest_path_vs_rewiring_T10000}).
Together with the clustering coefficient, the shortest-path length is a defining feature of small-world networks.
Although evidence of small-world properties of functional networks obtained from fMRI recordings have been provided in several studies (\eg~\citep{VandenHeuvel2008}), whether or not the brain is a small-world network is still being debated~\citep{Hilgetag2015,Papo2016}.
Following~\citet{Papo2016}, the question addressed here is of a pragmatic rather than an ontological nature: independently of whether the brain is a small-world network or not, to what extent can neuroscientists using standard system-level neuroimaging techniques interpret the small-world construct in the context of functional brain networks?
An indication that the interpretation is problematic was provided by~\citet{Hlinka2012}, who showed that functional connectivity matrices of randomly coupled autoregressive processes show small-world properties.
The effect is due to intrinsic properties of correlation rather than just to the finite sample size problem or spatial oversampling.
Specifically, correlation has a transitivity property: for any node $X$ with neighbours $Y$ and $Z$ (and respective correlations $\rho_{XY}$ and $\rho_{XZ}$), a lower bound can be derived for the correlation between the neighbours~\citep{Langford2001}:
\begin{equation} \label{eq:correlation_lower_bound}
    \rho_{YZ} \geq \rho_{XY}\rho_{XZ}-\sqrt{1-\rho_{XY}^2}\sqrt{1-\rho_{XZ}^2}.
\end{equation}
In particular, a strong positive correlation between two pairs of them implies a positive correlation within the third pair: $\rho^2_{XY}+\rho^2_{XZ}>1$ implies $\rho^2_{YZ}>0$~\citep{Langford2001}.
The problem was further investigated by~\citet{Zalesky2012}, who showed that functional connectivity matrices of independent processes also exhibit small-world properties and that---in practice---the correlation between neighbours is much higher than the theoretical lower bound in \eq{correlation_lower_bound}.
These considerations on correlation extend to bivariate MI, given the one-to-one relationship between MI and the absolute value of Pearson's correlation coefficient for the Gaussian variables considered in this study (see \ifarXiv\secRef{methods_bMI}\fi). 
This transitivity property results in more triangular cliques in functional networks, \ie an inflated clustering coefficient across the whole spectrum of networks in  \fig{WS_clustering_vs_rewiring_T10000}.
Together with the underestimate of the shortest-path length discussed above, the outcome is an overestimate of the small-worldness coefficient (\fig{WS_SW_coeff_vs_rewiring_T10000}).
As shown, the limitations of bivariate methods can be overcome by multivariate TE, to a large degree for shorter time series and certainly when sufficiently long time series are available.

\section{Scale-free networks}
\label{sec:BA}

\subsection{Numerical simulations}
The linear preferential attachment algorithm without attractiveness~\citep{Barabasi1999} is used to generate undirected scale-free networks of $N=200$ nodes.
Starting with two connected nodes, a new node is added at each iteration and linked bidirectionally to two existing nodes, selected with probability proportional to their current degree (via linear preferential attachment).
This preferential mechanism makes high-degree nodes more likely to be selected and further increase their degree---a positive feedback loop that generates few highly-connected hubs and many low-degree nodes.
The resulting density is approximately $4/N=0.02$ (average in- and out-degrees being approximately $4$), with hubs having degrees up to around \num{50} here.
A constant uniform link weight $C_{XY}=C_{XX}=0.1$ is assigned to all the links, achieving strong coupling but ensuring the stationarity of the VAR dynamics.
For robustness, each simulation is repeated \num{10} times on different network realisations and with random initial conditions.

\subsection{Results}

\begin{figure}
    \ifarXiv\includegraphics[width=0.48\textwidth]{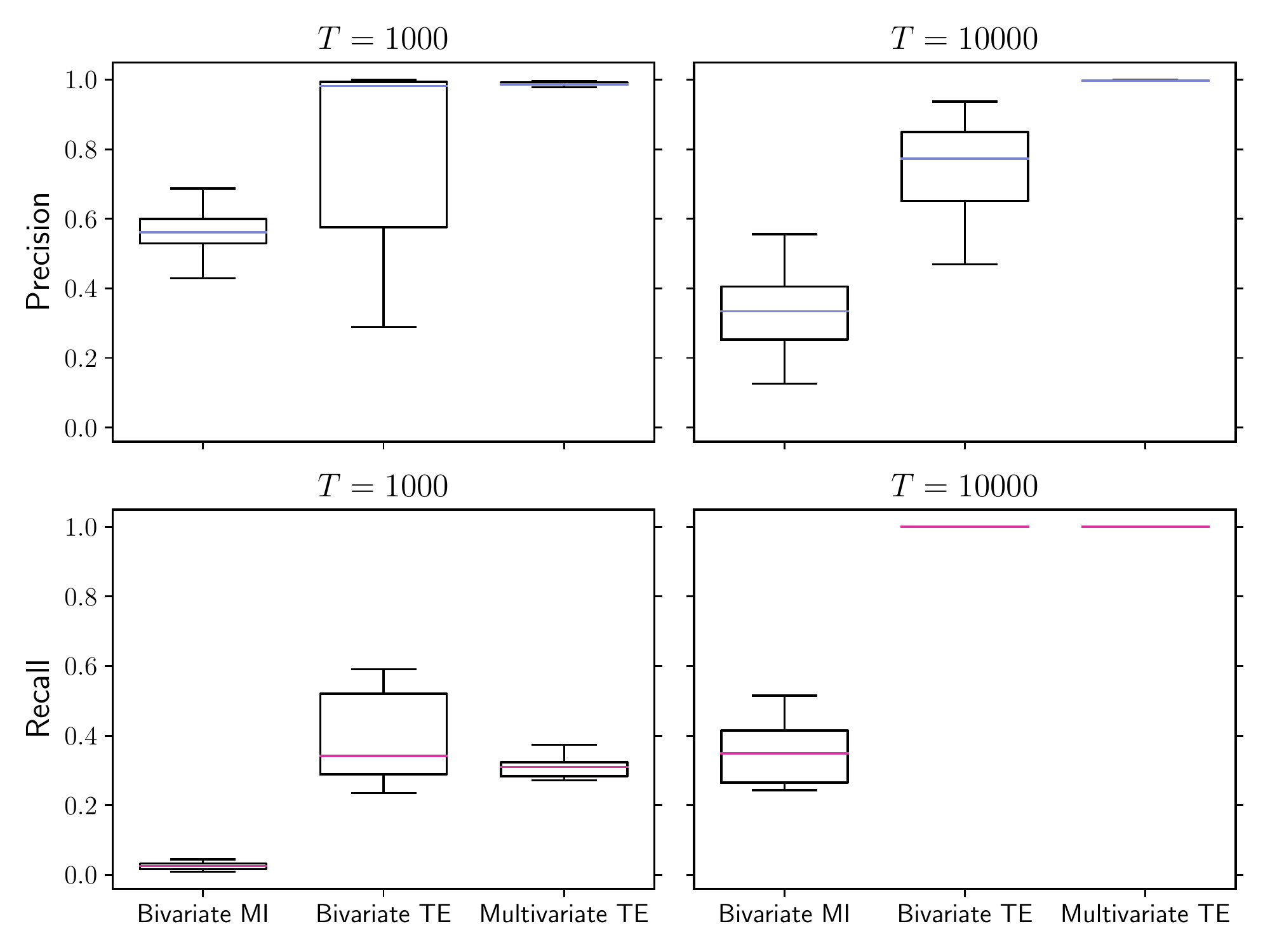}
    \fi
    \caption{\label{fig:BA_m2_performance_boxplot}
        Precision (top row) and recall (bottom row) in scale-free networks obtained via preferential attachment ($N$=\num{200} nodes).
        Multivariate TE guarantees high precision regardless of the amount of data available ($T=1000$ in the left column and $T$=\num{10000} in the right column).
        Bivariate TE can achieve slightly better recall than multivariate TE for shorter time series (bottom left panel) but its precision drops substantially for longer time series (top right panel) and the optimal time series length cannot be determined a priori.
        Bivariate MI has lower precision and recall than TE-based methods, both for $T$=\num{1000} and $T$=\num{10000}.
        The box-whiskers plots summarise the results over~\num{10} simulations on different network realisations, with median values indicated in colour.
    }
\end{figure}

At the microscale, the performance is evaluated in terms of precision and recall in the classification of the links.
The outcome is qualitatively similar to the small-world case presented above:
for longer time series (\num{10000} samples), multivariate TE is the most accurate method, achieving optimal performance according to all metrics (\fig{BA_m2_performance_boxplot}, right column).
Bivariate TE also achieves optimal recall; however, despite the strict statistical significance level, the precision is significantly lower than multivariate TE.
The bivariate MI algorithm scores comparatively very poorly both in terms of precision and recall ($<40$\% on average).
As expected, the recall of all methods decreases when shorter time series of \num{1000} samples are provided (\fig{BA_m2_performance_boxplot}, left column).
Once more, bivariate TE attains better precision on shorter time series than on longer ones, and for these networks attains slightly better recall than multivariate TE on the shorter time series.

\begin{figure}
    \ifarXiv\includegraphics[width=0.48\textwidth]{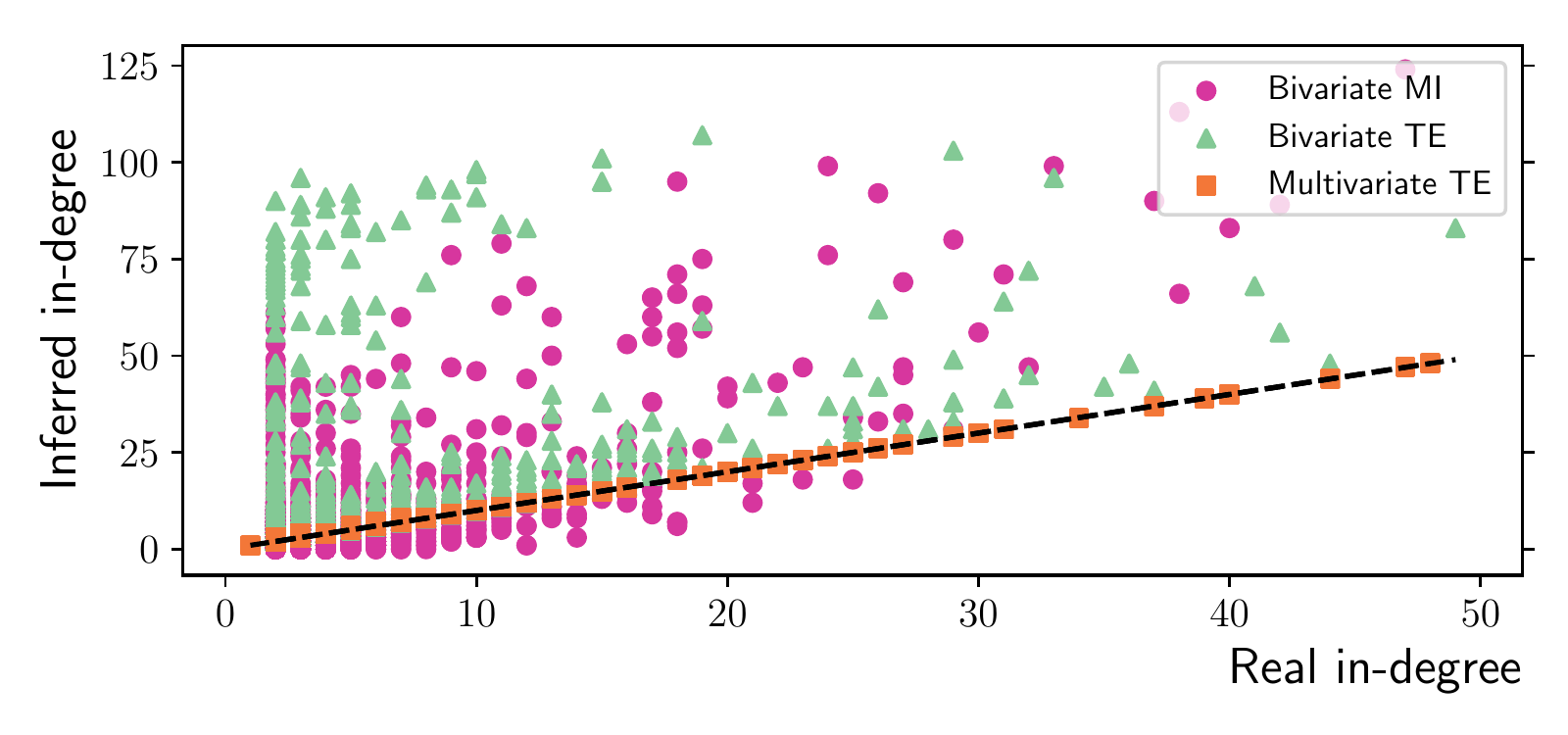}
    \fi
    \caption{\label{fig:BA_m2_in_degree}
        Inferred vs. real in-degree in scale-free networks obtained via preferential attachment ($N$=\num{200} nodes and $T$=\num{10000} time samples).
        Multivariate TE is the only algorithm able to preserve the in-degrees of the nodes as compared to their value in the real networks (ground truth).
        The dashed black line represents the identity between real and inferred values.
        Surprisingly, bivariate methods can inflate the in-degree of non-hubs by over one order of magnitude, making hubs less distinguishable.
        The results are collected over \num{10} simulations on different network realisations. 
    }
\end{figure}

\begin{figure}
    \ifarXiv\includegraphics[width=0.48\textwidth]{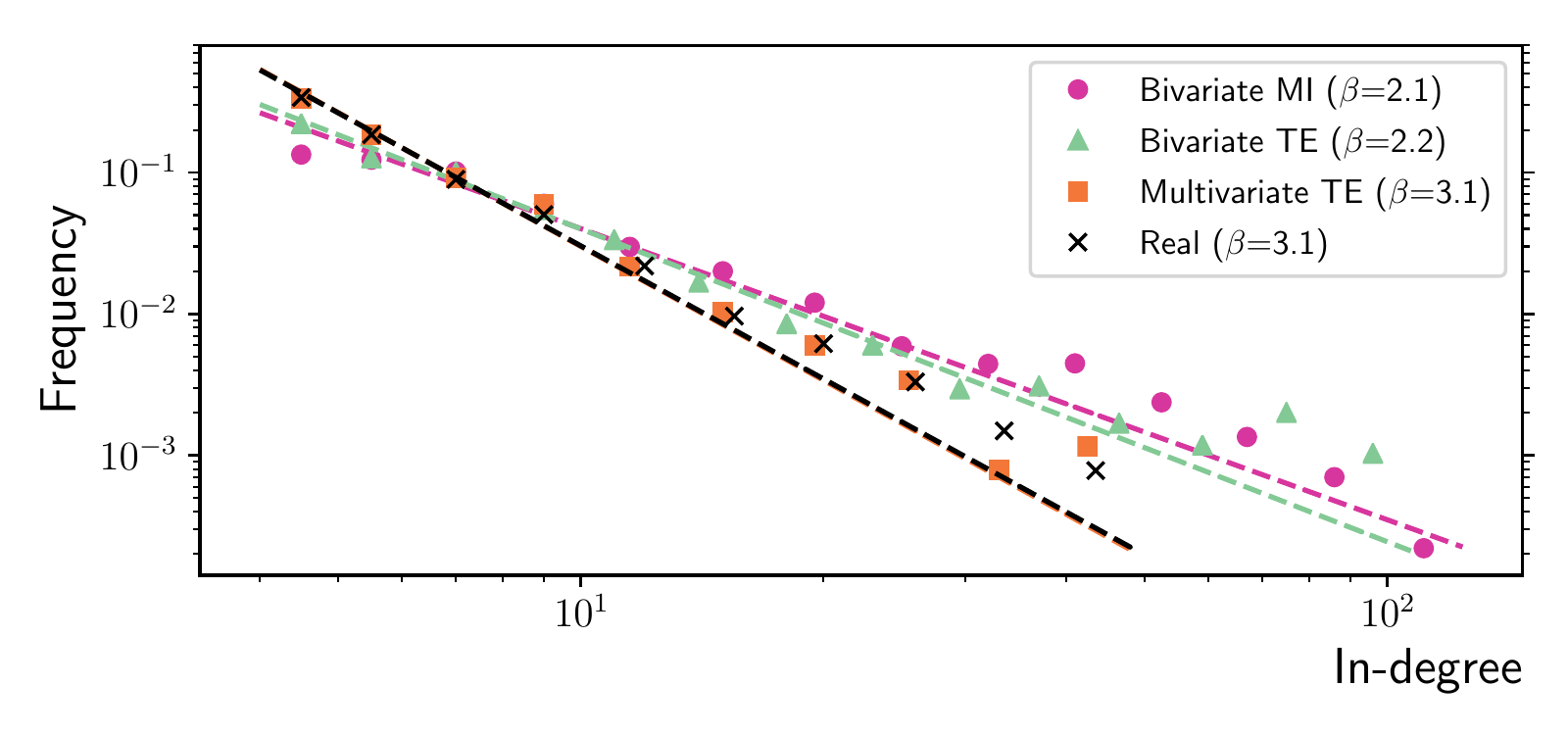}
    \fi
    \caption{\label{fig:BA_m2_in_degree_distr_loglog}
        Log-log plot of the in-degree distribution in scale-free networks obtained via preferential attachment ($N$=\num{200} nodes and $T$=\num{10000} time samples).
        The best power-law distribution fit for the real networks (ground truth) and the inferred network models are plotted with dashed lines (decay exponents reported in the legend).
        Despite the finite-size effect due to the small network size, multivariate TE is able to approximate the theoretical power-law decay exponent $\beta=3$ and to match the power-law fit to the real in-degree distribution ($\beta=3.1$).
        On the other hand, bivariate TE and MI underestimate the absolute value of the exponent.
        The results are collected over \num{10} simulations on different network realisations. 
        }
\end{figure}

At the macroscale, the three algorithms are tested on their ability to accurately measure several relevant properties of scale-free networks.
It is well known that the degree distribution of networks generated via this preferential attachment algorithm follows a power-law, with theoretical exponent $\beta=3$ in the limit of large networks~\citep{Barabasi1999}.
Fitting power-laws to empirical data requires some caution, \eg adopting a logarithmic binning scheme~\citep{Virkar2014}, and the dedicated \textit{powerlaw} Python package is employed for this purpose~\citep{Alstott2014}.
For sufficiently long time series ($T$ = \num{10000} in this study), multivariate TE is able to accurately recover the in-degrees of the nodes in our scale-free networks, while the bivariate MI and TE algorithms produce significant overestimates (\fig{BA_m2_in_degree}).
As a consequence, the (absolute value of the) exponent of the fitted power-law is underestimated by the latter methods, as shown in \fig{BA_m2_in_degree_distr_loglog}.

\begin{figure}
    \ifarXiv\includegraphics[width=0.48\textwidth]{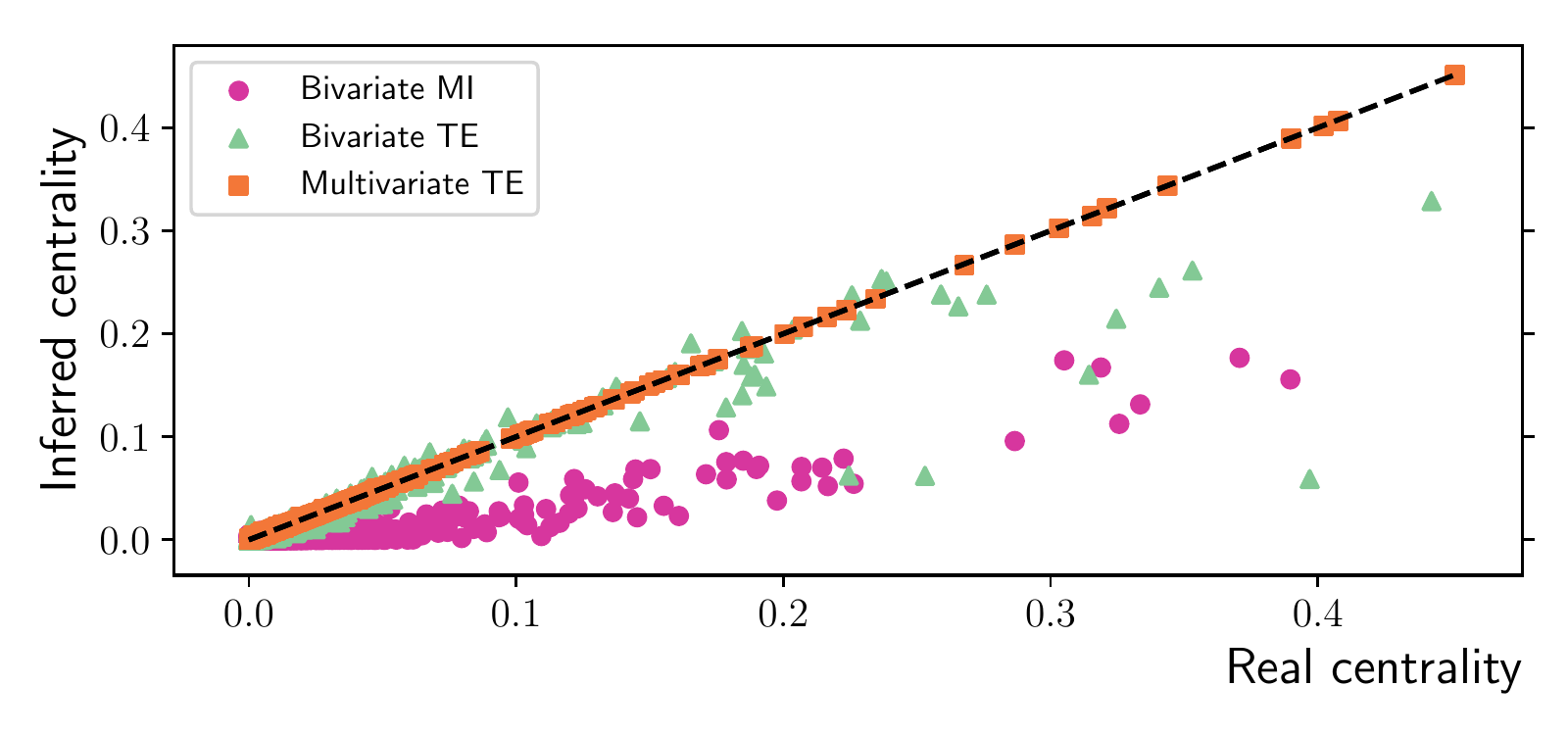}
    \fi
    \caption{\label{fig:BA_m2_centrality_betweenness}
        Inferred vs. real betweenness centrality for nodes in scale-free networks obtained via preferential attachment ($N$=\num{200} nodes and $T$=\num{10000} time samples).
        Multivariate TE is the only algorithm able to preserve the centrality of the nodes as compared to their value in the real networks (ground truth).
        The dashed black line represents the identity between real and inferred values.
        Bivariate MI underestimates the centrality of all nodes, while bivariate TE particularly underestimates the centrality of the most central nodes in the ground-truth network.
        The results are collected over \num{10} simulations on different network realisations. 
    }
\end{figure}

Hubs are a key feature of scale-free networks and have high betweenness centrality, since most shortest paths pass through them.
However, their centrality is highly underestimated by bivariate methods, often making it indistinguishable from the centrality of peripheral nodes~(\fig{BA_m2_centrality_betweenness}).

\begin{figure}
    \ifarXiv\includegraphics[width=0.48\textwidth]{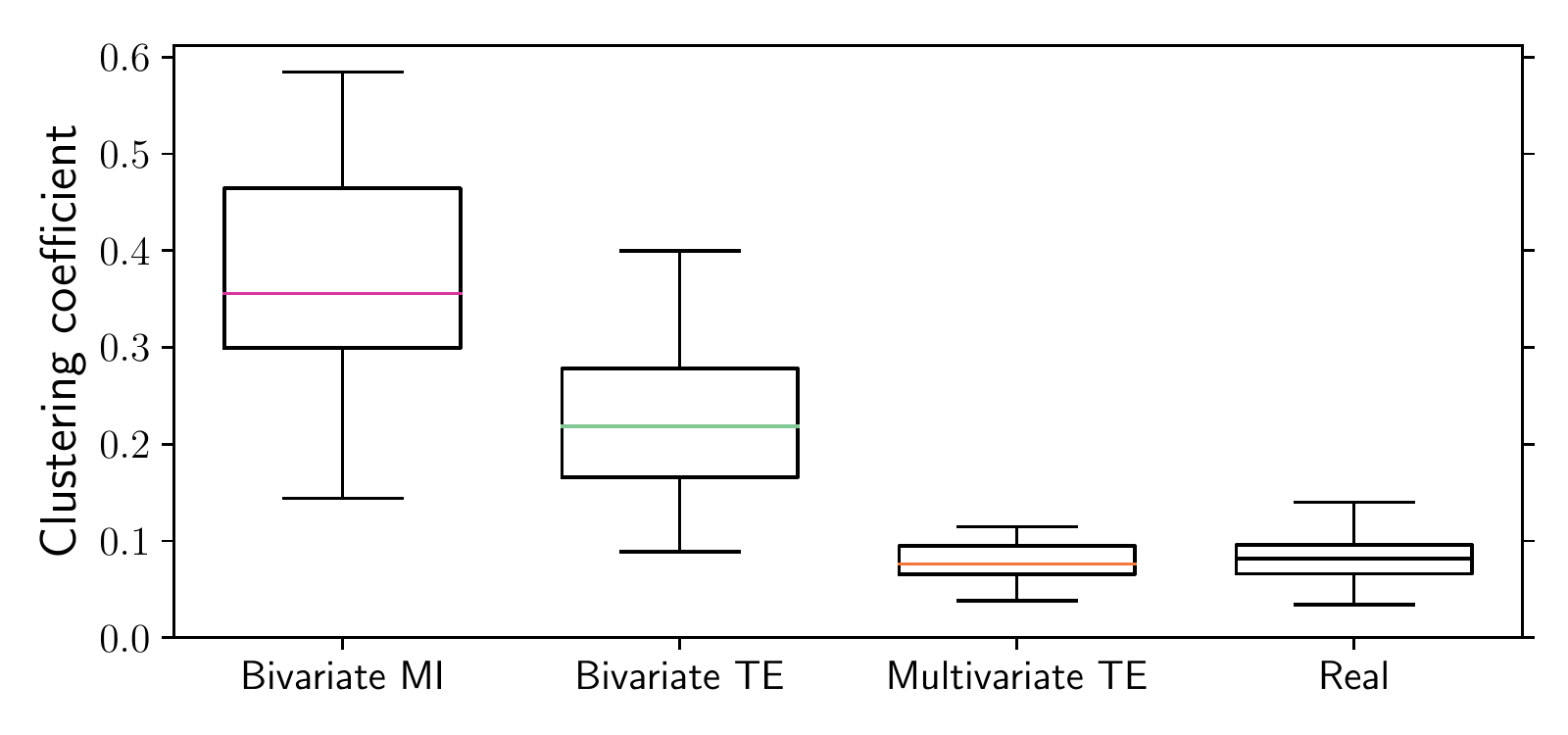}
    \fi
    \caption{\label{fig:BA_m2_clustering_boxplot}
        Inferred vs. real average clustering coefficient in scale-free networks obtained via preferential attachment ($N$=\num{200} nodes and $T$=\num{10000} time samples).
        Multivariate TE is the only algorithm able to preserve the average clustering of the real networks (ground truth), while bivariate TE and MI consistently overestimate it.
        The box-whiskers plots summarise the results over~\num{10} simulations, with median values indicated in colour.
    }
\end{figure}

As in the small-world case, multivariate TE is able to very closely approximate the real mean clustering coefficient, while the bivariate MI and TE algorithms consistently overestimate it (\fig{BA_m2_clustering_boxplot}).
The related measure of local efficiency~\citep{Latora2001} is reported in \ifarXiv\appRef{efficiency}\fi.
A closer examination of the clustering of individual nodes (instead of the average) reveals that low clustering values are consistently overestimated by bivariate methods
, while high clustering values are underestimated (\ifarXiv\appRef{BA_clustering}\fi).
Finally, bivariate methods overestimate the rich-club coefficient (\ifarXiv\appRef{BA_rich_club}\fi).

\subsection{Discussion}
Echoing the discussion of small-world networks \ifarXiv in \secRef{WS}\fi, the ability to control the false positives while building connectomes---exhibited only by multivariate TE---is also crucial for correctly identifying fundamental features of scale-free networks, such as the power-law degree distribution and the presence of hub nodes.
Hubs are characterised by high degree and betweenness centrality.
Unfortunately, the centrality of hubs is not robust with respect to false positives: the addition of spurious links cause strong underestimates of the betweenness centrality of real hubs, since additional links provide alternative shortest paths.
For bivariate TE, the effect is so prominent that the inferred centrality of real hubs can be indistinguishable from the centrality of peripheral nodes, as shown in \fig{BA_m2_centrality_betweenness}.
The in-degree is in principle more robust with respect to false positives; however, bivariate methods infer so many spurious incoming links into non-hubs that they become as connected (or more) than the real hubs are inferred to be (\fig{BA_m2_in_degree}).
Taken together, these effects on the in-degree and centrality greatly hinder the identification of real hubs when bivariate MI or TE are employed.
The inflation of the in-degree of peripheral nodes also fattens the tail of the in-degree distribution (\fig{BA_m2_in_degree_distr_loglog}), resulting in an underestimate of the exponent of the fitted power-law with respect to the theoretical value $\beta=3$~\citep{Barabasi1999}.
This has severe implications for the synthetic networks used in this study, erroneously providing evidence against the simple preferential attachment algorithm used to generate them.
The third distinct characteristic of these networks is their low average clustering, which is also induced by the preferential attachment algorithm, whereby each new node is only connected to two existing ones.
However, bivariate methods fail to capture this feature, producing a strong overestimate of the average clustering coefficient (\fig{BA_m2_clustering_boxplot}).
This can be attributed to the transitivity property of Pearson's correlation, which produces overabundant triangular cliques in functional networks (as previously discussed).
Given the significant biases affecting all the distinctive properties of scale-free networks---in addition to the small-world networks presented above---it is evident that great caution should be used when applying bivariate inference methods (cross-correlation, MI, TE) to draw conclusions as to topological properties of real-world networks.
In contrast, again, the multivariate TE was demonstrated to produce network models with microscopic and macroscopic topological properties consistent with those of the underlying structural scale-free networks.

\section{Modular networks}
\label{sec:modular}

\subsection{Numerical simulations}
In order to study the performance of the three inference algorithms on modular topologies, networks of \num{100} nodes are generated and equally partitioned into five groups of \num{20}.
Initially, each node is directly linked to \num{10} random targets within its own group, such that the five communities are completely disconnected.
The initial density is thus \num{50}\% within each group and \num{10}\% overall.
Link targets are then gradually rewired from within to between groups, weakening the modular structure but preserving the overall density and keeping the out-degrees fixed.
Eventually, the concepts of ``within'' and ``between'' groups are no longer meaningful---the links are equally distributed and the topology resembles a random Erd\H{o}s-R\'{e}nyi network of equal overall density.
This happens when the rewiring is so prevalent that only \num{2} links are left within the initial groups and \num{8} out of \num{10} links are formed between them (for each node).
Going even further, when all \num{10} links are formed between the initial groups and none within, the network becomes multipartite, \ie the nodes are partitioned into five independent sets having no internal connections.
A constant uniform link weight $C_{XY}=C_{XX}=0.08$ is assigned to all the links, achieving strong coupling but ensuring the stationarity of the VAR dynamics.
Each simulation is repeated \num{10} times on different network realisations and with random initial conditions.

\subsection{Results}

\begin{figure}
    \ifarXiv\includegraphics[width=0.48\textwidth]{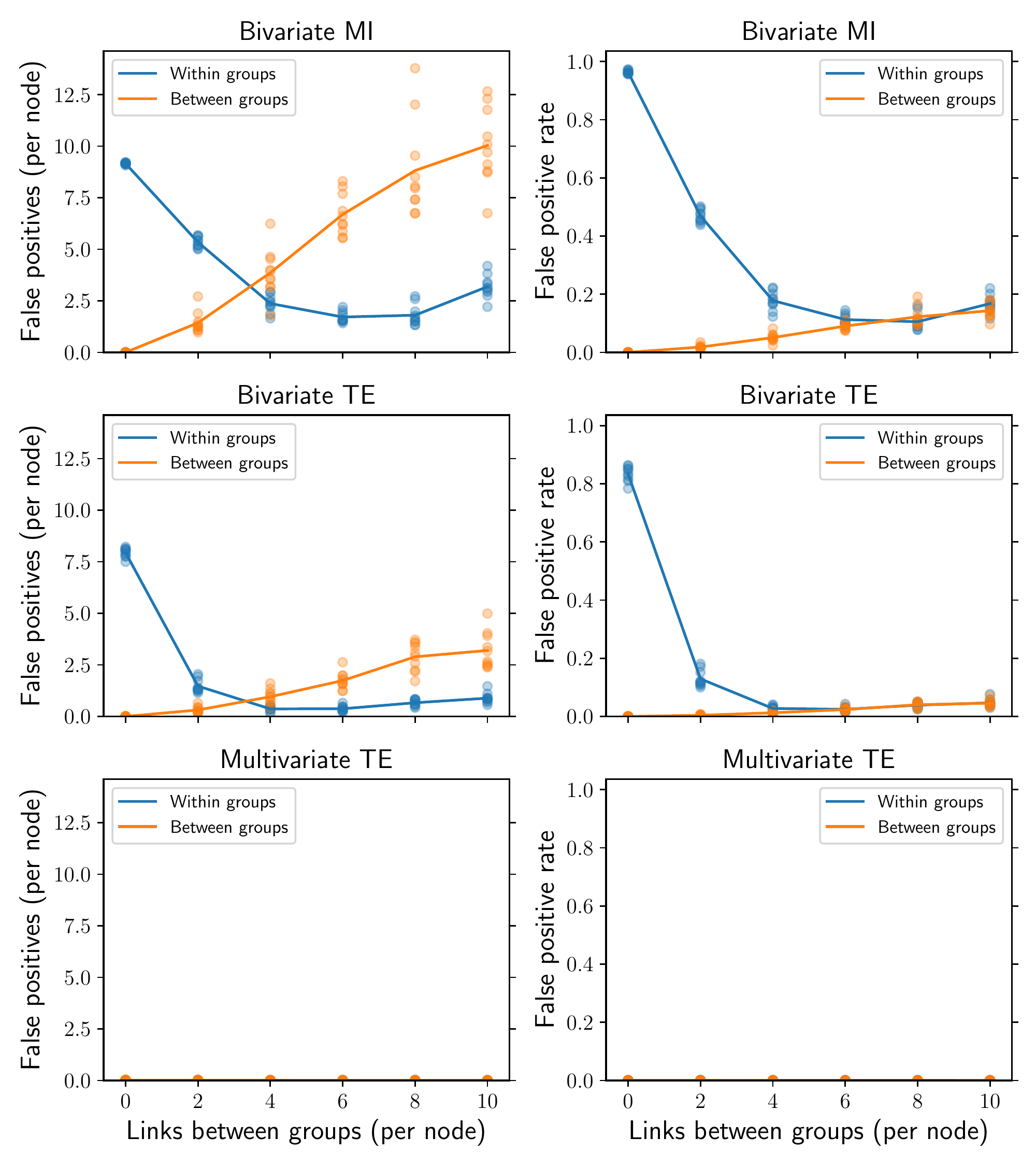}
    \fi
    \caption{\label{fig:modular_T10000_FP_and_FP_rate_vs_links_out}
        False positives per node (left column) and false-positive rate (right column) in modular networks of $N$=\num{100} nodes and $T$=\num{10000} time samples.
        Each node is connected to \num{10} neighbours and the horizontal axis represents the number of links between groups (in the two extreme cases all \num{10} links are formed exclusively within each group or exclusively between groups).
        Bivariate MI and TE infer more spurious links (left column) within the initial groups than between them for smaller between-group densities; the comparison then reverses for larger between-group densities.
        The false-positive rate (\ie the normalised number of false positives; right column) is also higher within groups for smaller between-group densities, whilst the rate for between groups becomes comparable (instead of larger) as between-group density increases.
        The number of false positives produced by multivariate TE is comparatively negligible.
        The results for~\num{10} simulations on different networks are presented (low-opacity markers) in addition to the mean values (solid markers).
    }
\end{figure}

At the microscale, we find that bivariate MI and TE infer more spurious links within the initial groups than between them for smaller between-group densities (\fig{modular_T10000_FP_and_FP_rate_vs_links_out}, left column).
As the between-group density increases though, we find more spurious links between the initial groups than within them.
The normalised false-positive rate is also significantly higher within groups for smaller between-group densities (right column), however the normalisation sees the false-positive rate becoming comparable between and within group as the between-group density increases.
The number of false positives produced by multivariate TE is comparatively negligible.

\begin{figure}
    \ifarXiv\includegraphics[width=0.48\textwidth]{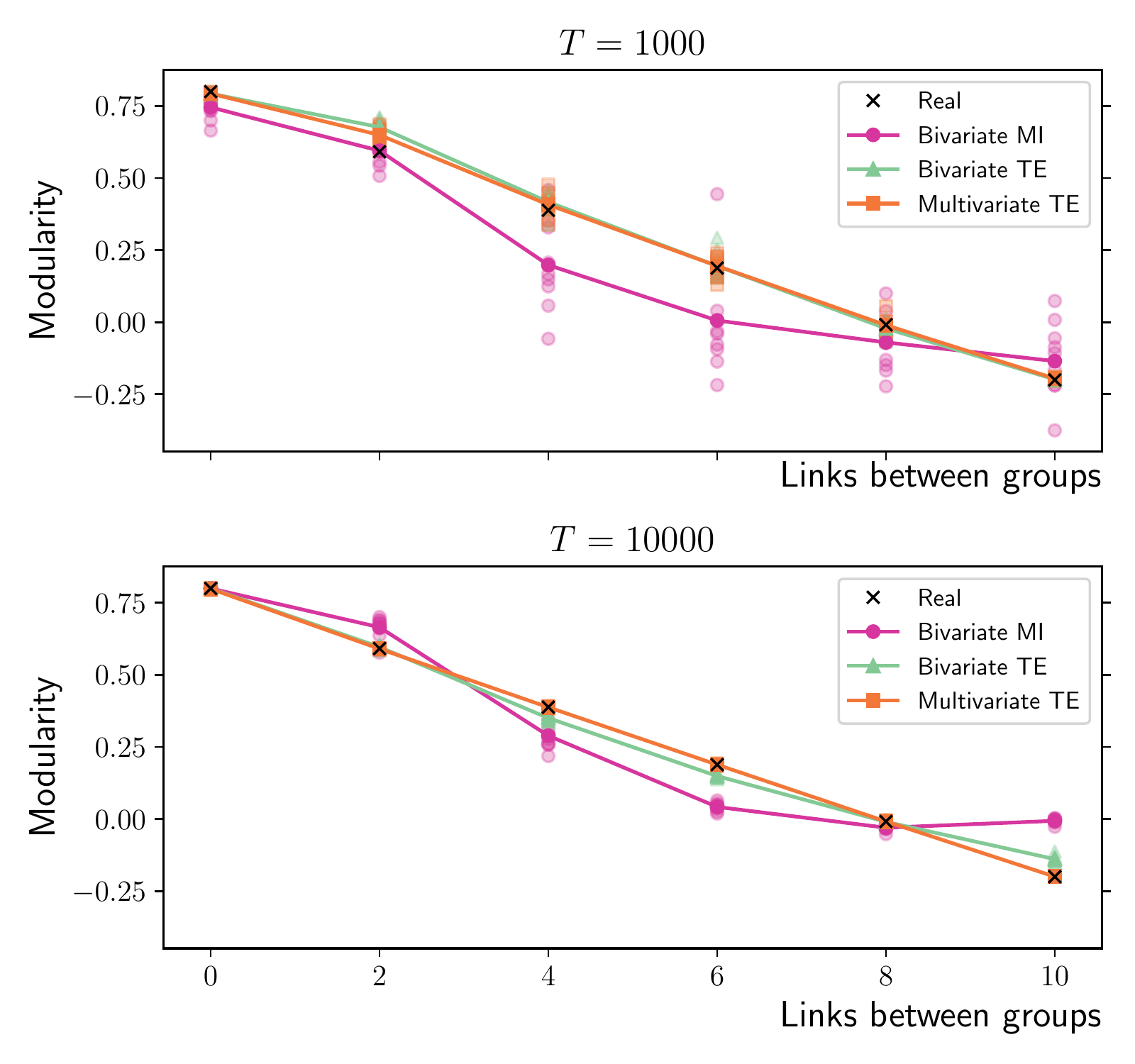}
    \fi
    \caption{\label{fig:modularity_vs_links_out}
        Modularity of the partition representing five groups of \num{20} nodes ($T$=\num{10000} time samples).
        Each node is initially connected to \num{10} neighbours within the same group via directed links.
        As link targets are gradually rewired from within to between groups, the modularity of the initial partition linearly decreases.
        The horizontal axis represents the number of links between groups for each node (in the two extreme cases all \num{10} links are formed exclusively within each group or exclusively between groups).
        Bivariate and multivariate TE produce accurate estimates of the real modularity, while bivariate MI often underestimates it, both for shorter and longer time series ($T$=\num{1000} in the top panel and $T$=\num{10000} in the bottom one).
        The results for~\num{10} simulations on different networks are presented (low-opacity markers) in addition to the mean values (solid markers).
    }
\end{figure}

At the mesoscale, the modularity of the partition corresponding to the five disconnected communities is maximal in the absence of rewiring and decreases as more and more links are formed between groups rather than within them (\fig{modularity_vs_links_out}).
Bivariate and multivariate TE produce accurate estimates of the real modularity, while bivariate MI often underestimates it, particularly for shorter time series ($T$=\num{1000}) and intermediate between-group densities.

\subsection{Discussion}
Our results on modular networks confirm and extend previous findings on correlation-based functional connectivity, stating that ``\textit{false positives occur more prevalently between network modules than within them, and the spurious inter-modular connections have a dramatic impact on network topology}''~\citep{Zalesky2016}.
Indeed, the left column of \fig{modular_T10000_FP_and_FP_rate_vs_links_out} shows that bivariate MI and TE infer a larger \textit{number} of false positives between the initial groups than within them, once we have a mid-range between-group link density in the underlying structure (which induces the transitive relationships).
However, the same does not apply to the false positive \textit{rate} (\ie the normalised number of false positives) shown in the right column of \fig{modular_T10000_FP_and_FP_rate_vs_links_out}: where an edge does not actually exist, it is more likely to be inferred if it is within rather than across group (for up to mid-range between-group link densities).
As such, the higher number of false positives between modules is mostly due to the larger number of potential spurious links available between different communities compared to those within them.
Nonetheless, the key message is that the modular structure (at the mesoscale level) affects the performance of bivariate algorithms in inferring single links (at the microscale level).
This provides further empirical evidence for the theoretical finding that bivariate TE---despite being a pairwise measure---does not depend solely on the directed link weight between a single pair of nodes, but on the larger network structure they are embedded in, via the mesoscopic network motifs~\citep{Novelli2020}.
In particular, the abundance of specific ``clustered motifs'' in modular structure increase the bivariate TE, making links within each group easier to detect but also increasing the false-positive rate within modules.
Other studies have related also the correlation-based functional connectivity to specific structural features, such as search information, path transitivity~\citep{Goni2014}, and topological similarity~\citep{Bettinardi2017}.

The underestimate of the modularity of the initial partition by bivariate MI (\fig{modularity_vs_links_out}, bottom panel) is a direct result of these higher numbers of spurious between-group links.
This has important implications for the identification of the modules, since a lower score makes this partition less likely to be deemed optimal by popular greedy modularity maximisation algorithms~\citep{Blondel2008}.
We speculate that the spurious inter-modular links would also hinder the identification of the modules when alternative approaches for community detection are employed (a thorough comparison of which is beyond the scope of this study).

\section{Macaque connectome}
\label{sec:macaque}

Finally, the three inference algorithms are compared on two real macaque brain connectomes, using both linear VAR dynamics and a nonlinear neural mass model (pipeline illustrated in \fig{pipeline}).

\subsection{Numerical simulations}
\subsubsection{Linear VAR dynamics}
As a final validation study under idealised conditions, the linear VAR dynamics in \eq{VAR} is run on the connectome obtained via tract-tracing by \citet{Young1993}.
This directed network consists of \num{71} nodes and \num{746} links ($15\%$ density) and incorporates multiple properties investigated in the previous sections, including a small-world topology and the presence of hubs and modules.
The scaling of the performance is studied as a function of the cross-coupling strength (\ie the sum of incoming link weights into each node, denoted as $C_\textup{in}$ and formally defined as $C_\textup{in}=\sum_X C_{XY}$ for each node $Y$).
The coupling $C_\textup{in}$ is varied in the $[0.3,0.7]$ range, making $C_{XY}$ constant for each parent $X$ for a given $Y$ to achieve this, and the self-link weights are kept constant at $C_{XX}=0.2$ to ensure the stationarity of the VAR dynamics.
For robustness, each simulation is repeated \num{10} times with random initial conditions.

\subsubsection{Nonlinear neural mass model}
As a final experiment, we provide an initial investigation of whether the insights from the previous validation studies extend beyond the idealised conditions there.
Specifically, in moving towards a more realistic setting, neural mass model dynamics are simulated on the CoCoMac connectome, as described in \ifarXiv\secRef{methods}\fi.
This network structure contains $76$ nodes with $1560$ directed connections ($27\%$ density), which are weighted and have experimentally estimated coupling delays.
Importantly, by incorporating nonlinear coupling, coupling delays, a distribution of coupling weights, and subsampling, this last study drops many of the simplifying assumptions made using the VAR dynamics in the previous sections.
The linear Gaussian estimator is retained for our information-theoretic measures despite the nonlinear interactions here, so as to remain consistent with the previous studies.
Dropping the assumption of sampling at the real causal process resolution adds a particular challenge, and is often encountered in practice in modalities with low temporal resolution.
To handle the variation in coupling delays, we consider sources at lags up to $L=4$ time steps (60 ms) here.
The longest time-series analysed (\num{30000} samples) corresponds to \num{7.5} minutes of one sample per \num{15} ms.

\subsection{Results}
\begin{figure}
    \ifarXiv\includegraphics[width=0.48\textwidth]{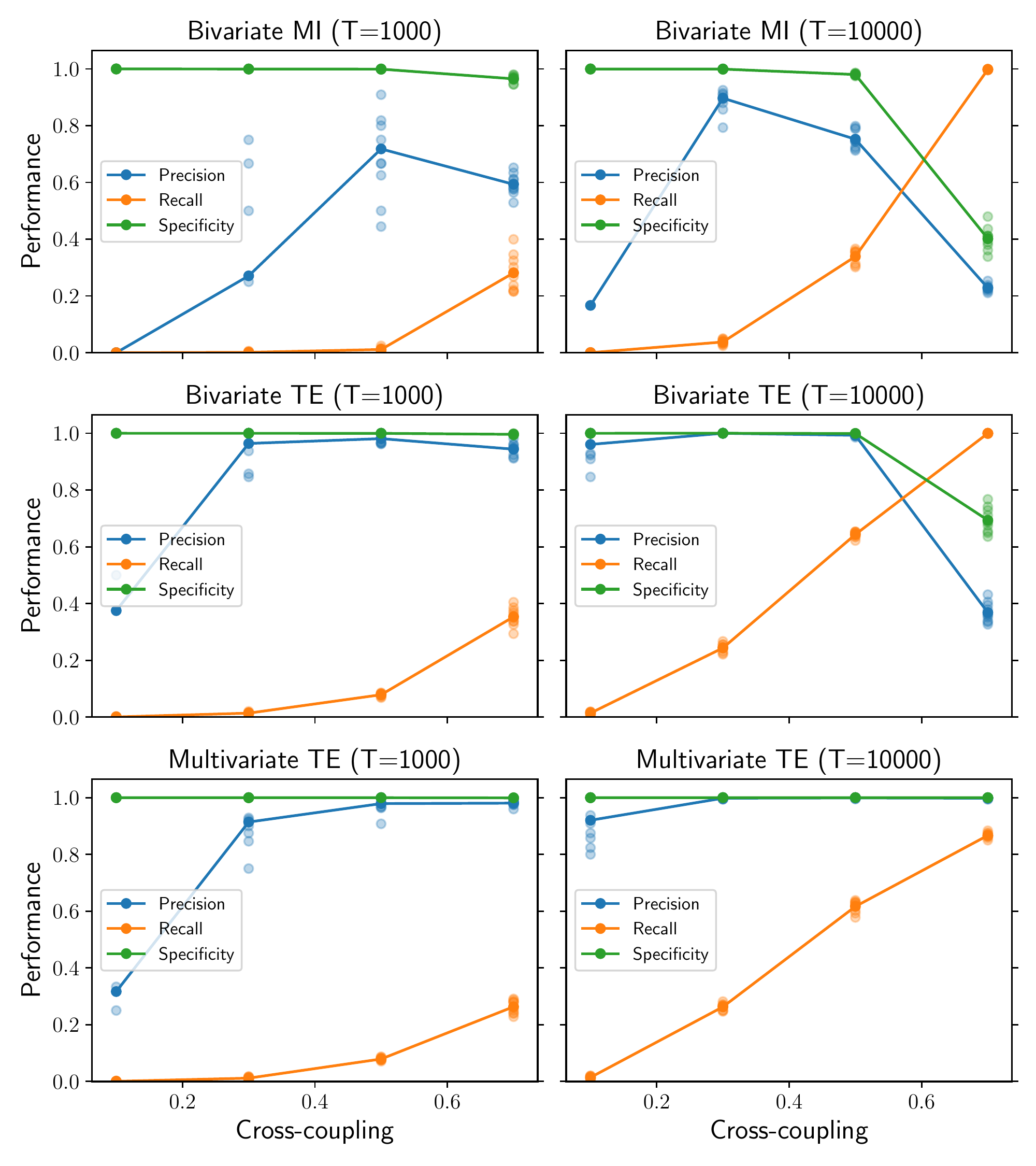}
    \fi
    \caption{\label{fig:macaque71_performance_vs_total_cross_coupling}
        Performance as a function of coupling weight in a real macacque connectome with \num{71} nodes and linear VAR dynamics.
        Multivariate TE (third row) guarantees high specificity (adequate control of the false-positive rate) regardless of the cross-coupling strength and time series length ($T$=\num{1000} in the left column and and $T$=\num{10000} in the right column).
        Bivariate TE (second row) attains similar performance to multivariate TE for low cross-coupling strength and short time series, according to all metrics.
        For stronger coupling or longer time series, the recall of all methods increase, but the precision and specificity of the bivariate methods substantially drop.
        For each value of the cross-coupling weights, the results for~\num{10} simulations from random initial conditions are presented (low-opacity markers) in addition to the mean values (solid markers).
    }
\end{figure}

\begin{figure}
    \ifarXiv\includegraphics[width=0.48\textwidth]{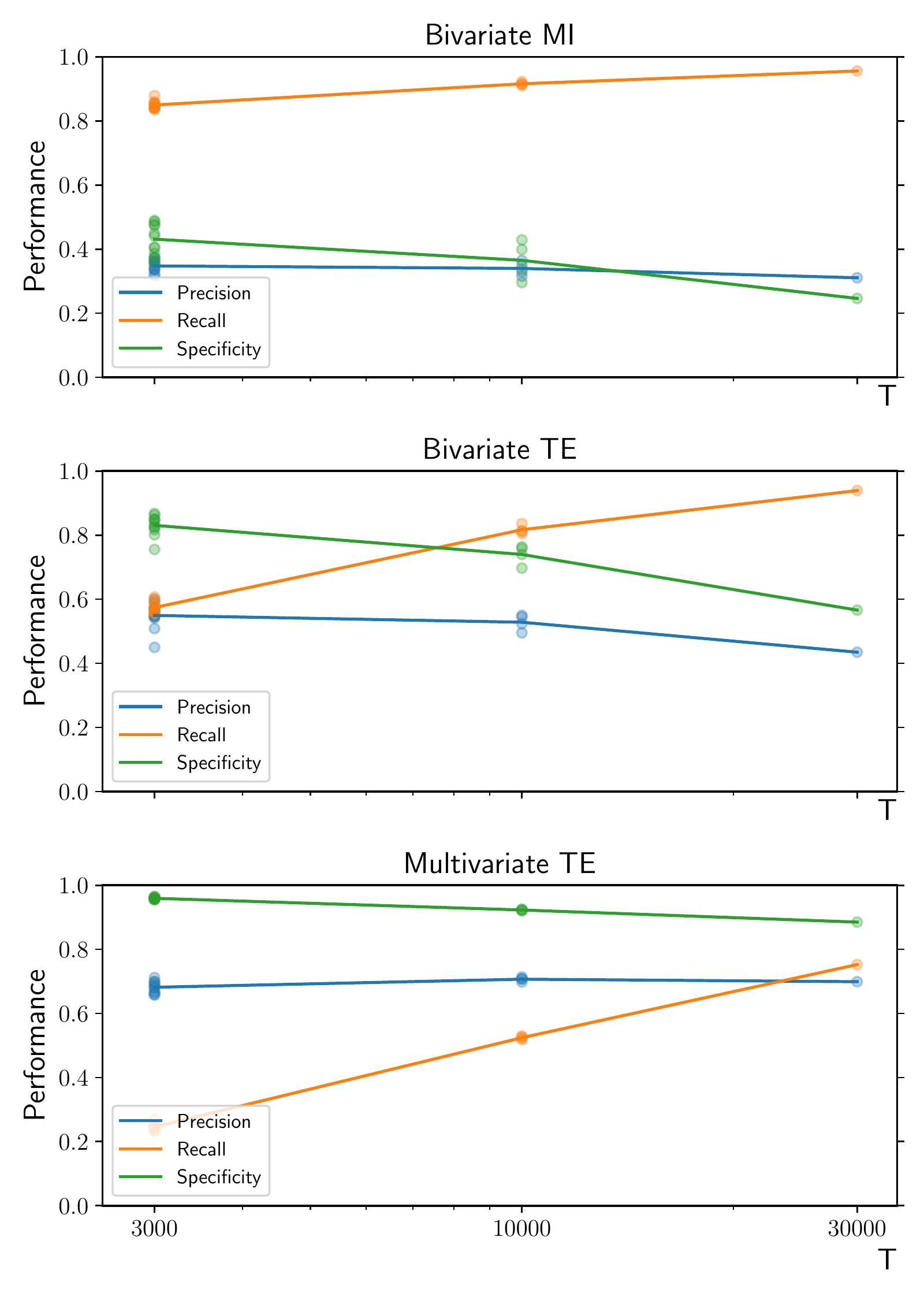}
    \fi
    \caption{\label{fig:macaque76_performance_vs_T}
        Performance as a function of number of time series samples ($T$) in the CoCoMac connectome with \num{76} nodes and nonlinear dynamics (neural mass model).
        Similarly to the linear VAR dynamics case presented in the previous figure, multivariate TE (third row) has lower recall than bivariate methods, particularly for shorter time series.
        More importantly though, it guarantees higher and more consistent precision and specificity, testifying to a more effective control of false positives.
        This advantage becomes increasingly important as longer time series are provided.
        The results for all simulations are presented (low-opacity markers) in addition to the mean values (solid markers).
    }
\end{figure}

At the microscale, the results for the linear and nonlinear dynamics (\fig{macaque71_performance_vs_total_cross_coupling} and \fig{macaque76_performance_vs_T}) are complementary and summarise the main findings presented so far.
There exists a window---characterised by low cross-coupling strength and short time series---where bivariate TE attains similar or better performance compared to multivariate TE in terms of recall, specificity, and precision.
For stronger coupling or longer time series, the recall of all methods increase, but the precision and specificity of the bivariate methods substantially drop whilst those of multivariate TE remain consistently high.

\begin{figure*}
    \ifarXiv\includegraphics[width=\textwidth]{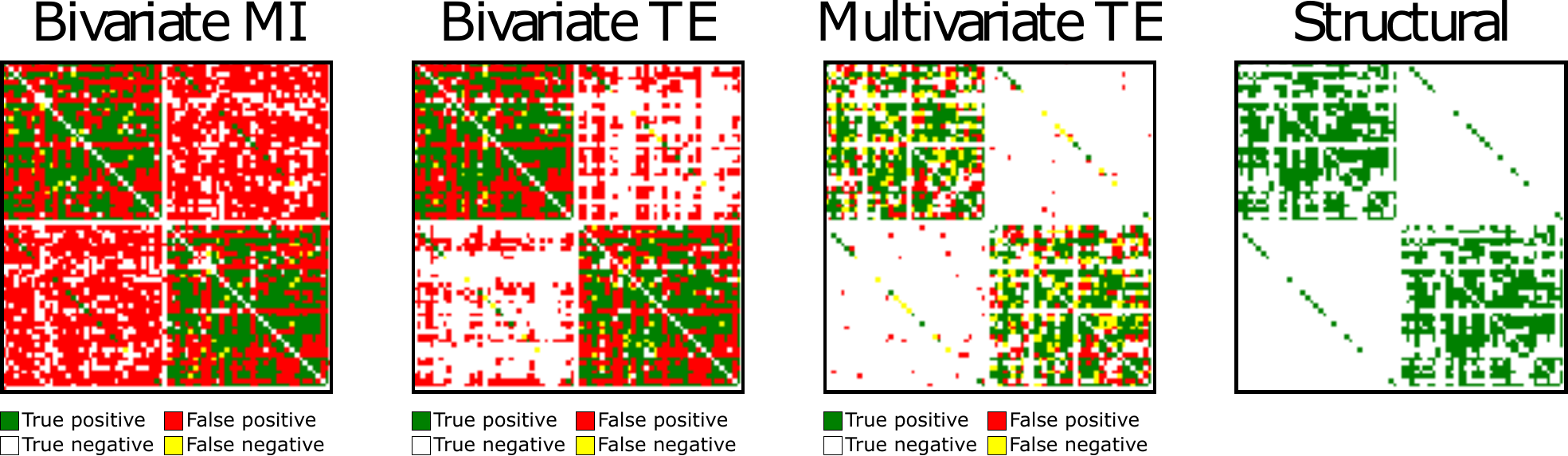}
    \fi
    \caption{\label{fig:macaque76_30000samples_adj_mats}
        Inferred adjacency matrices beside real  CoCoMac connectome with $N$=\num{76} nodes.
        Rows are source regions, columns are targets.
        Different colours indicate which links are correctly/incorrectly inferred or missed by each inference method.
        True positives are indicated in green, false positives in red, false negatives in yellow, and true negatives in white.
        The large number of spurious inter-hemispheric links produced by bivariate methods can hinder the identification of important macroscopic features, including the two hemispheres (particularly for longer time series, as in the case of $T=30000$ time samples shown here).
    }
\end{figure*}

An intuitive visual representation of how these differences in precision, recall, and specificity affect the macroscopic inferred network is provided in \fig{macaque76_30000samples_adj_mats}, where the inferred adjacency matrices are displayed beside the real connectome, with different colours indicating which links are correctly/incorrectly inferred or missed by each method.
The macroscale results (in terms of local and global efficiency measures) are reported in \ifarXiv\appRef{efficiency}\fi.

\subsection{Discussion}
Interestingly, \fig{macaque71_performance_vs_total_cross_coupling} and \fig{macaque76_performance_vs_T} show how similar outcomes are produced by either stronger coupling (link weights) or longer time series.
An explanation is readily available in the simple case of VAR dynamics: the bivariate TE on spurious links is typically lower than the TE on real links, and it increases with the coupling strength \ifarXiv\footnote{Bivariate TE can be analytically derived from the network structure under the assumption of VAR dynamics, and spurious links are associated with larger motifs (\eg longer chains), contributing to the TE at lower orders of magnitude~\citep{Novelli2020}.}.\fi
Therefore, spurious links can only pass statistical significance tests when sufficiently long time series are available (in order for their weak TE values to be distinguished from noise); for the same reason, for shorter time series, spurious links can only be detected in the presence of strong enough coupling. 
Unfortunately, for real datasets, there is no consistent a priori way to determine the optimal window of time-series lengths for bivariate TE, before increasing false-positive rate decays precision and specificity.

It is crucial to note that, despite moving beyond the idealised conditions used for validation with the VAR model, 
the qualitative differences between the inference algorithms remain unchanged in the neural mass model study.
That is, multivariate TE attains higher and more consistent precision and specificity than bivariate methods, testifying to a more effective control of false positives that enables more faithful representation of macroscale network features---an advantage that becomes increasingly important as longer time series are provided.
Indeed, on a more intuitive level, \fig{macaque76_30000samples_adj_mats} provides immediate visual evidence of the importance of a reliable method for controlling the false-positive rate.
The large number of spurious inter-hemispheric links produced by bivariate methods can hinder the identification of important macroscopic features, starting from the very presence of the two hemispheres themselves and extending to other fundamental network properties (as shown for local and global efficiency measures in \ifarXiv\appRef{efficiency}\fi).
This issue becomes particularly problematic for longer time series, as in the case of $T=30000$ time samples shown in \fig{macaque76_30000samples_adj_mats}.

With that said, the precision and specificity of the multivariate TE for this more realistic study in \fig{macaque76_performance_vs_T} is noticeably lower compared to the previous experiments in idealised conditions, such as that shown in \fig{macaque71_performance_vs_total_cross_coupling}.
Specifically, the specificity is lower than would be expected from the proven well-controlled false-positive rate under idealised conditions.
This is potentially due to a number of factors in this study, including the subsampling, nonlinear dynamics, strong autocorrelation and to a lesser extent coupling delays.
Whilst we retained the linear Gaussian estimator to be consistent with the previous experiments, we have previously demonstrated substantial performance enhancements for the multivariate TE algorithm when using a nonlinear estimator for studying nonlinear dynamics~\citep{Novelli2019}.
That could be expected to improve performance here as well.
Similarly, we have recently demonstrated approaches to rigorously control the known inflation of false-positive rates for MI and TE due to autocorrelation~\citep{Cliff2020}, although this is not expected to have as dramatic an effect in this study due to the selection of subsampling time via the autocorrelation time.
The subsampling itself though (by a factor of $30$ on the original time series) is likely to have had a significant impact on performance.
This is because subsampling obscures our view of the dynamics at the real interaction scale, and prevents us from properly conditioning on the past of the target (which is known to inflate false-positive rate \citep{Wibral2011}).
Whilst initial studies have suggested some level of robustness of TE to subsampling~\citep{Lizier2011}, a more systematic analysis would be important future work to properly understand its effect.

\section{Conclusion}
We have sought to evaluate how well network models produced via bivariate and multivariate network inference methods capture features of underlying structural topologies.
As outlined in the Introduction, these inference techniques seek to infer a network \textit{model} of the relationships between the nodes in a system, and are not necessarily designed nor expected to replicate the underlying structural topology \textit{in general}.
Our primary focus, however, was in evaluating the techniques under specific idealised conditions under which effective network models are proven to converge to the underlying structure.
The focus on such conditions is important because they provide assumptions under which our evaluation becomes a validation study.
The performance of these methods was evaluated at both the microscopic and macroscopic scales of the network.
Whilst we may not expect the same performance in identifying links at the microscopic scale,
we should expect all of the methods to identify relevant macroscopic features in the underlying network structure, as well as distinctive nodes or groups of nodes.

For longer time series, multivariate TE performs better on all network topologies (lattice-like, small-world, scale-free, modular, and the real macaque connectome).
This enhanced performance is very clear at the microscale of single links, achieving high precision and recall, and consequently at the macroscale of network properties, accurately reflecting the key summary statistics of the ground truth networks used for validation.

Bivariate methods (directed and undirected) can exhibit higher recall (or sensitivity) for shorter time series for certain underlying topologies; however, as available data increases, they are unable to control false positives (that is, they have lower specificity).
Whilst decreasing statistical significance thresholds (critical $\alpha$ levels) for inferring links is a common strategy to reduce false positives, the bivariate measures simply cannot match the sensitivity of the multivariate approach at the same specificity (compare the precisions for same recall level in \fig{BA_m2_performance_boxplot} as an example, or refer to the sample ROC curve in \ifarXiv\appRef{roc_curve}\fi~for a more extensive comparison).
At the macroscale, the comparatively larger number of false positives leads to overestimated clustering, small-world, and rich-club coefficients, underestimated shortest path lengths and hub centrality, and fattened degree distribution tails. 
The changes in these measures are partly due to the aforementioned transitivity property for bivariate measures (implying that false positives are often `close' to real links in the network), and partly due to higher density; untangling these effects is a topic for future work.
In any case, caution should therefore be used when \textit{interpreting} network properties of functional connectomes obtained via correlation or pairwise statistical dependence measures.
Their use is only advisable when the limited amount of data doesn't allow the use of the more sophisticated but more accurate multivariate TE, which more faithfully tracks trends in underlying structural topology.
Further research is required to try to reliably identify---\textit{a priori}---situations where bivariate TE will exhibit higher precision and recall (particularly in terms of time series length), as there is no clear candidate approach to do so at present.
In the current status quo, the critical strength of the multivariate approach lies in its ability to appropriately control the false-positive rate to meet the requested values.

Our evaluation of the inference techniques under idealised conditions considered several of the highest profile complex network topologies: lattice-like, small-world, scale-free, modular, and a mix of their features in a real macaque connectome.
This complements previous work \citep{Novelli2019,Sun2015} at a similar scale, which evaluated performance on random network structures and incorporated a study of the effect of network size and linear-vs-nonlinear dynamics and estimators.
Obviously, we have only scratched the surface of examining the effects of the myriad combinations of network parameters that could be investigated, which could include larger variations in degree, distributions on edge weights, super- or sub-linear preferential attachment, non-uniform module sizes or cross-module connection probabilities, and could also incorporate experiments across other types of dynamics.
Thus far, our conclusions on how the multivariate TE approach performs against the bivariate measures were consistent across the variety of structures, and whilst it would be interesting to see how other variations in structure effect the performance, we do expect the general conclusions on the comparison between approaches to remain similar.

Of course, there is a computational time trade-off, with the run-time of the multivariate TE algorithm requiring $\mathcal{O}(d)$ longer in comparison to the bivariate approach (where $d$ is the average inferred in-degree).
The run-time complexity is analysed in detail and benchmarked for both linear and nonlinear estimators on similar scale experiments in \citep[Supporting Information]{Novelli2019}.
Our experiments here on $10000$ time samples for up to $200$ nodes, with the more efficient linear estimator, took less than $2$ hours (single core) on average per target on the same hardware.
Given the availability of parallel computing to analyse targets simultaneously, we believe the trade-off in run-time increase is justifiable for the performance increase demonstrated here.

Beyond idealised conditions, effective network inference techniques are not guaranteed to converge in such manner to an underlying structure.
This can be for many reasons, including hidden nodes (or lack of full observability), non-stationarity or short sample size, or sub-sampling obscuring the scale of interaction.
Yet, our final experiment (examining time series dynamics of a neural mass model on the $76$ node CoCoMac connectome) extended our investigations into the domain beyond idealised conditions and also demonstrated superior performance of the multivariate TE, aligning with the validation studies in idealised conditions.
Importantly, this included visually revealing the characteristic hemispheric macroscopic structure of this connectome.
With that said, the performance of multivariate TE in this example was certainly reduced in comparison to our experiments under idealised conditions.
This appears to be due to various factors as discussed in that section, including the use of a linear estimator on nonlinear dynamics as well as the effect of subsampling.
There is substantial scope for further study to understand the performance of inference techniques under non-ideal techniques, and how the effective network models they infer are related to underlying structure.
This will involve further experiments on realistic neural dynamics; systematic study of the effect of subsampling in network inference (building on existing studies for Granger causality \citep{Barnett2017}), and assessing the ability of inference algorithms to capture key network features when only a subset of nodes is observed (\ie in the presence of hidden nodes).

Finally, while we focus on functional brain networks, our conclusions and methods also apply to anatomical brain networks in which connectivity is measured using correlation in cortical thickness or volume~\citep{He2007}.
Beyond neuroscience, they also extend to metabolite, protein and gene correlation networks~\citep{Gillis2011} (a similar validation study using synthetic networks was carried out in gene regulatory networks using bivariate MI and TE~\citep{Budden2016}).

\section*{Supporting Information}
The network inference algorithms described in this paper are implemented in the open-source Python software package IDTxl~\citep{Wollstadt2019}, which is freely available on GitHub (\url{https://github.com/pwollstadt/IDTxl}).
The code used for the systematic exploration of network structures and inference methods is also publicly available (\url{https://github.com/LNov/infonet}).

\ifarXiv
\begin{acknowledgments}
\else
\acknowledgments
\fi
JL was supported through the Australian Research Council DECRA Fellowship grant DE160100630 and through The University of Sydney Research Accelerator (SOAR) prize program.
The authors thank Mac Shine and Daniele Marinazzo for useful discussions.
The authors acknowledge the Sydney Informatics Hub and the University of Sydney's high-performance computing cluster Artemis for providing the high-performance computing resources that have contributed to the research results reported within this paper.
The authors thank Matthew Aburn for providing time-series data simulated from the neural mass model on the CoCoMac connectome from \cite{Shine2018,Li2019}.
\ifarXiv
\end{acknowledgments}
\else
\fi

\ifarXiv
\section*{Author Contributions}
\fi
Leonardo Novelli: Conceptualization; Data curation; Formal analysis; Investigation; Software; Validation; Visualization; Writing --- original draft.
Joseph T. Lizier: Conceptualization; Funding acquisition; Methodology; Supervision; Writing --- review \& editing.

\ifarXiv
\appendix
\fi

\begin{figure}[b]
    \ifarXiv\includegraphics[width=0.48\textwidth]{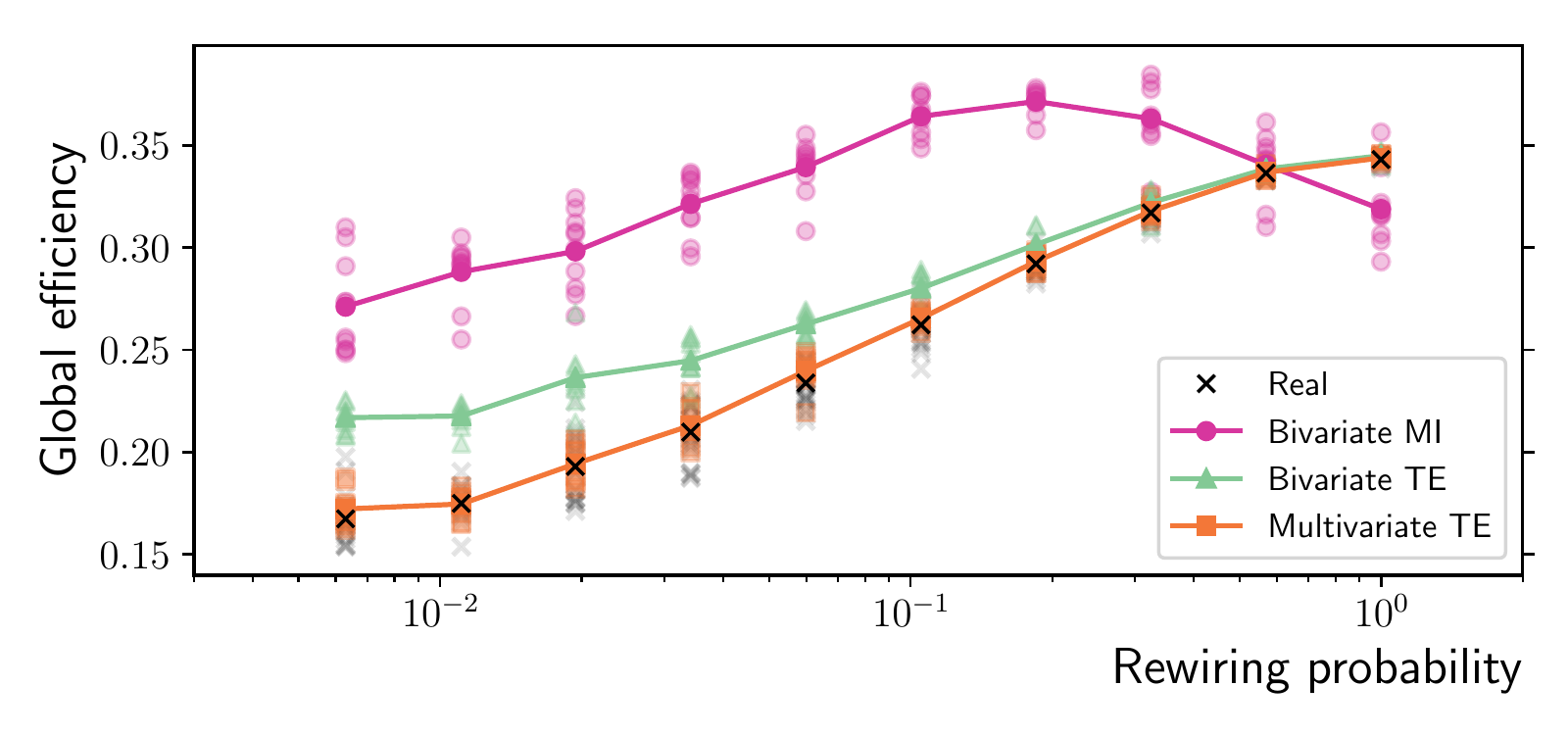}
    \fi

    \ifarXiv\includegraphics[width=0.48\textwidth]{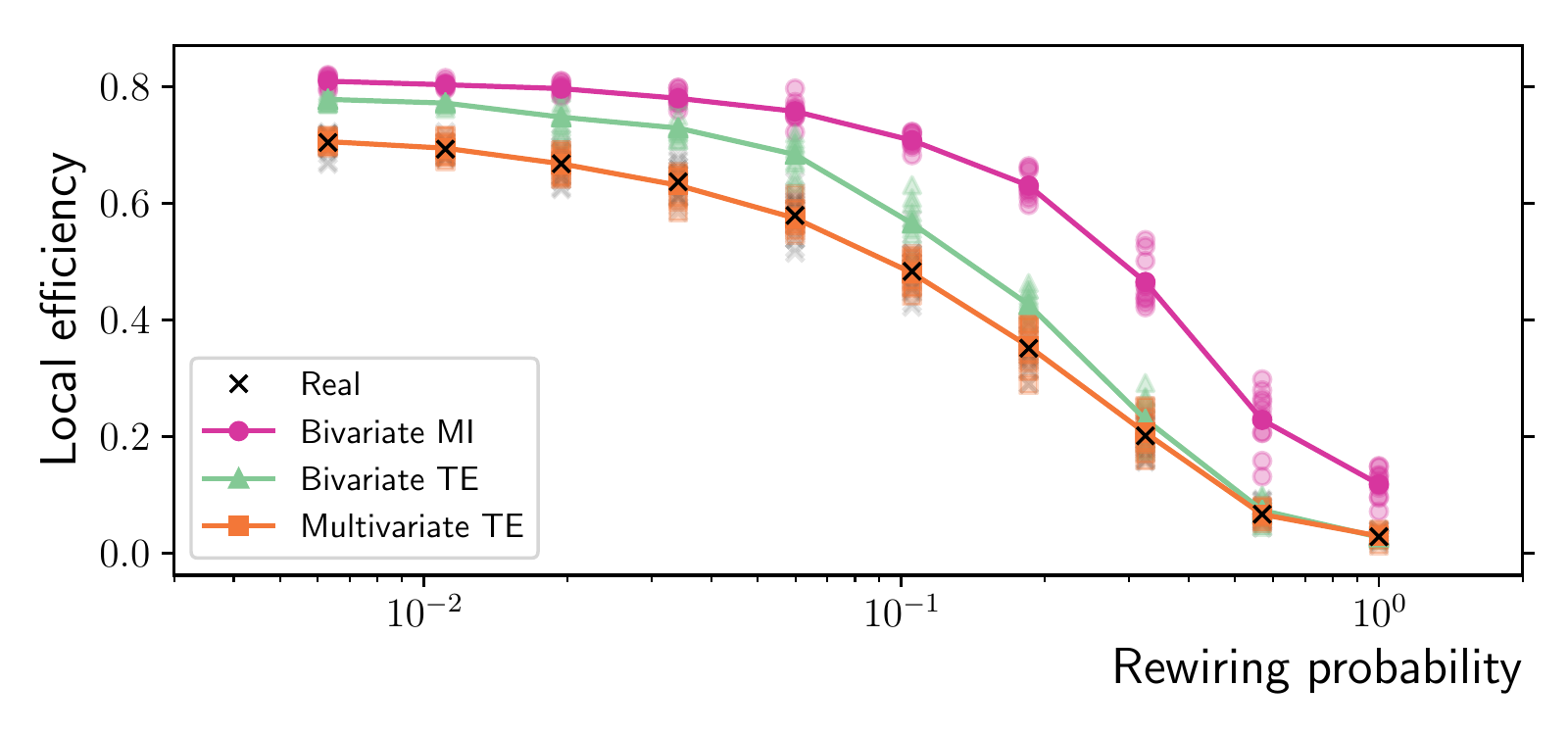}
    \fi
    \caption{\label{fig:WS_efficiency_vs_rewiring_T10000}
        Global and local efficiency as a function of the rewiring probability in Watts-Strogatz ring networks ($N$=\num{100} nodes and $T$=\num{10000} time samples).
        Multivariate TE reconstructs networks having the same efficiency as the real topologies (ground truth).
        On the other hand, bivariate MI and TE produce significant overestimates due to spurious links.
        These create shortcuts across the network (inflating the global efficiency in the top panel) and form spurious triangular cliques (inflating the local efficiency in the bottom panel), particularly on lattice-like topologies (low rewiring probability).
        For each value of the rewiring probability, the results for~\num{10} simulations on different network realisations are presented (low-opacity markers) in addition to the mean values (solid markers).
    }
\end{figure}

\begin{figure}
    \ifarXiv\includegraphics[width=0.45\textwidth]{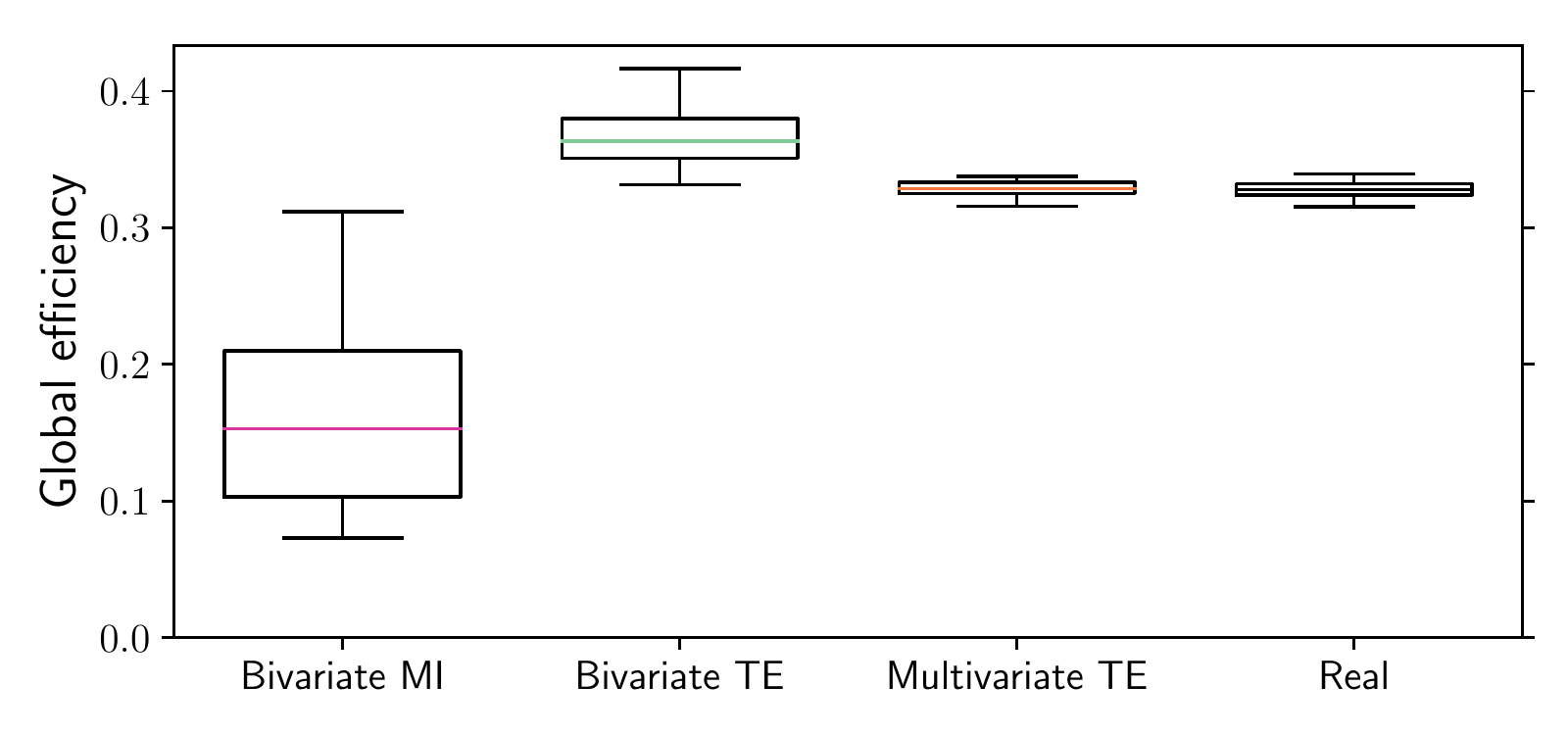}
    \fi
    
    \ifarXiv\includegraphics[width=0.45\textwidth]{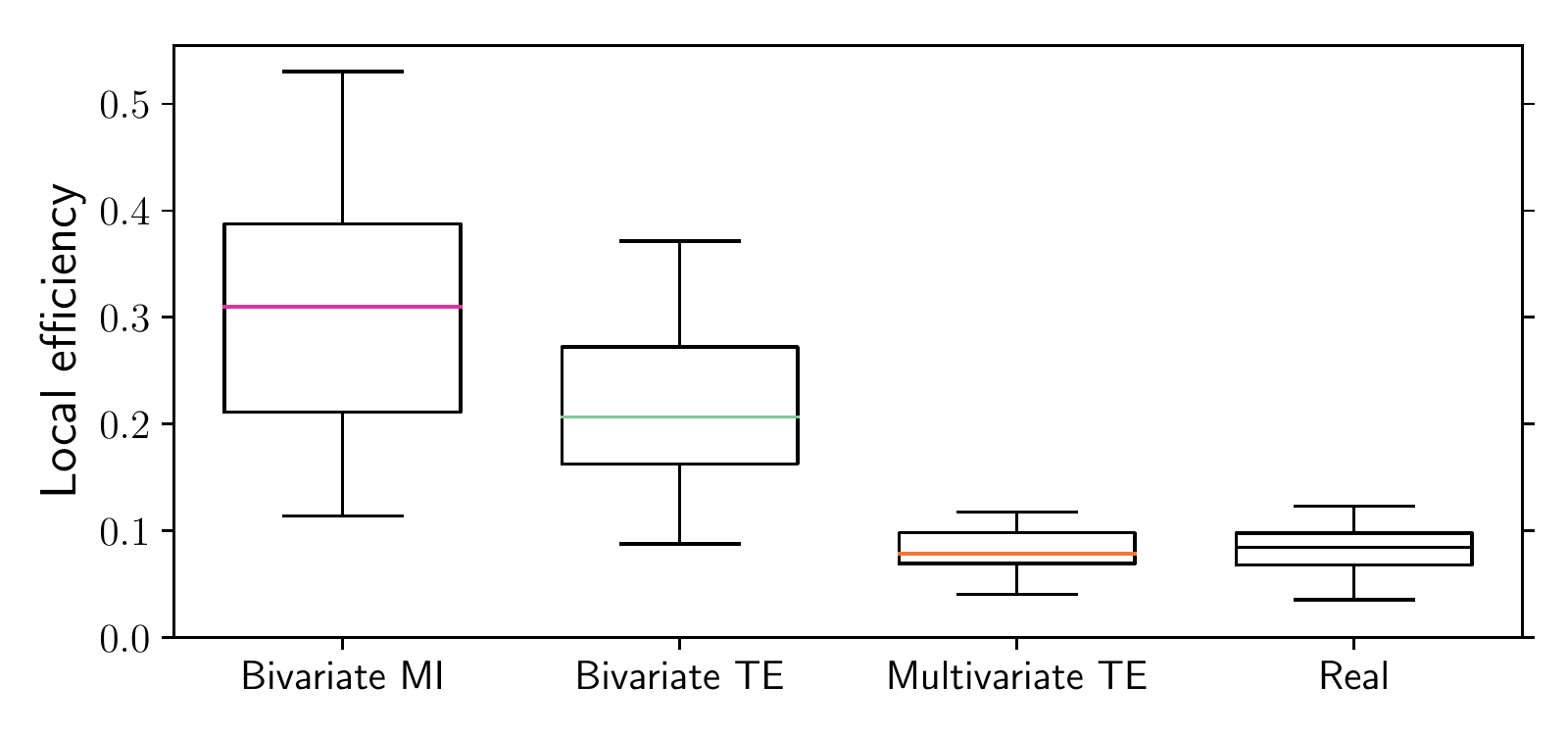}
    \fi
    \caption{\label{fig:BA_m2_efficiency}
        Global and local efficiency in scale-free networks obtained via preferential attachment ($N$=\num{200} nodes and $T$=\num{10000} time samples).
        Multivariate TE is the only algorithm able to preserve the efficiency of the real networks (ground truth), while bivariate TE and MI consistently overestimate it.
        The box-whiskers plots summarise the results over~\num{10} simulations, with median values indicated in colour.
    }
\end{figure}

\begin{figure}
    \ifarXiv\includegraphics[width=0.45\textwidth]{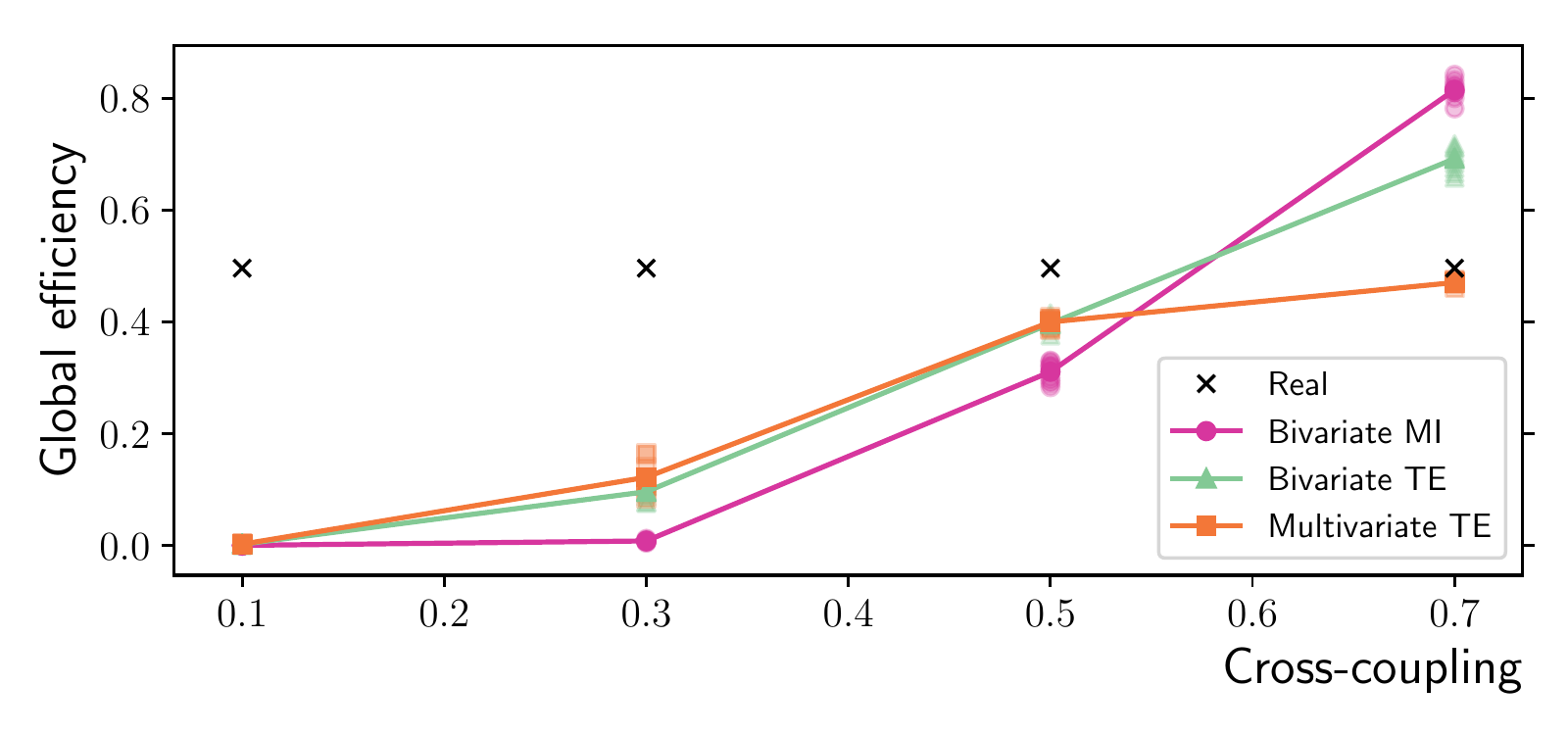}
    \fi
   
    \ifarXiv\includegraphics[width=0.45\textwidth]{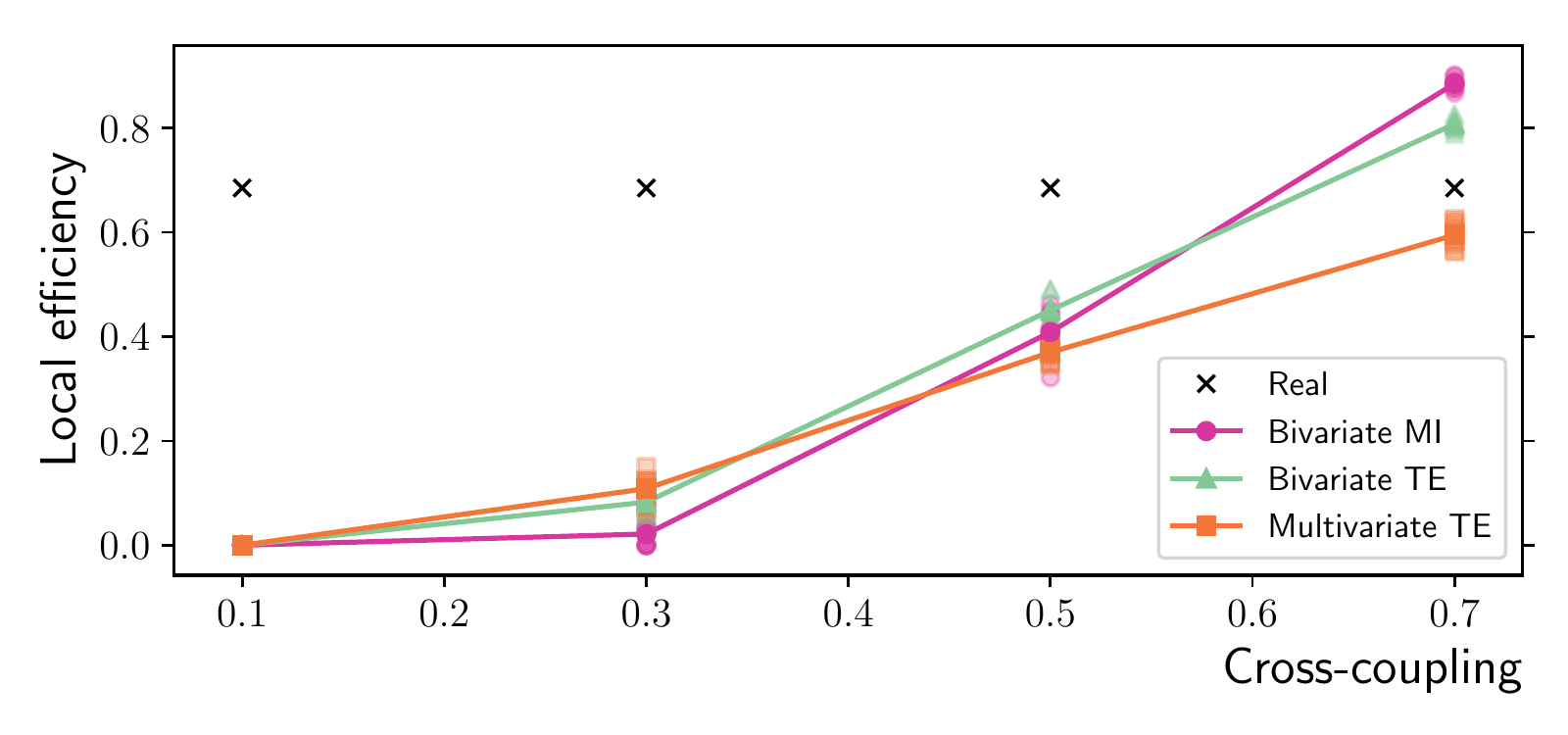}
    \fi
    \caption{\label{fig:macaque71_efficiency_vs_total_cross_coupling}
        Global and local efficiency as a function of coupling weight in a real macacque connectome ($N$=\num{71} nodes and $T$=\num{10000} time samples).
        All inference algorithms produce underestimates for low coupling.
        For stronger coupling, multivariate TE converges to the real global and local efficiency of the underlying networks (ground truth), while bivariate methods overestimate both measures.
        For each value of the cross-coupling weights, the results for~\num{10} simulations from random initial conditions are presented (low-opacity markers) in addition to the mean values (solid markers).
    }
\end{figure}

\begin{figure}
    \ifarXiv\includegraphics[width=0.45\textwidth]{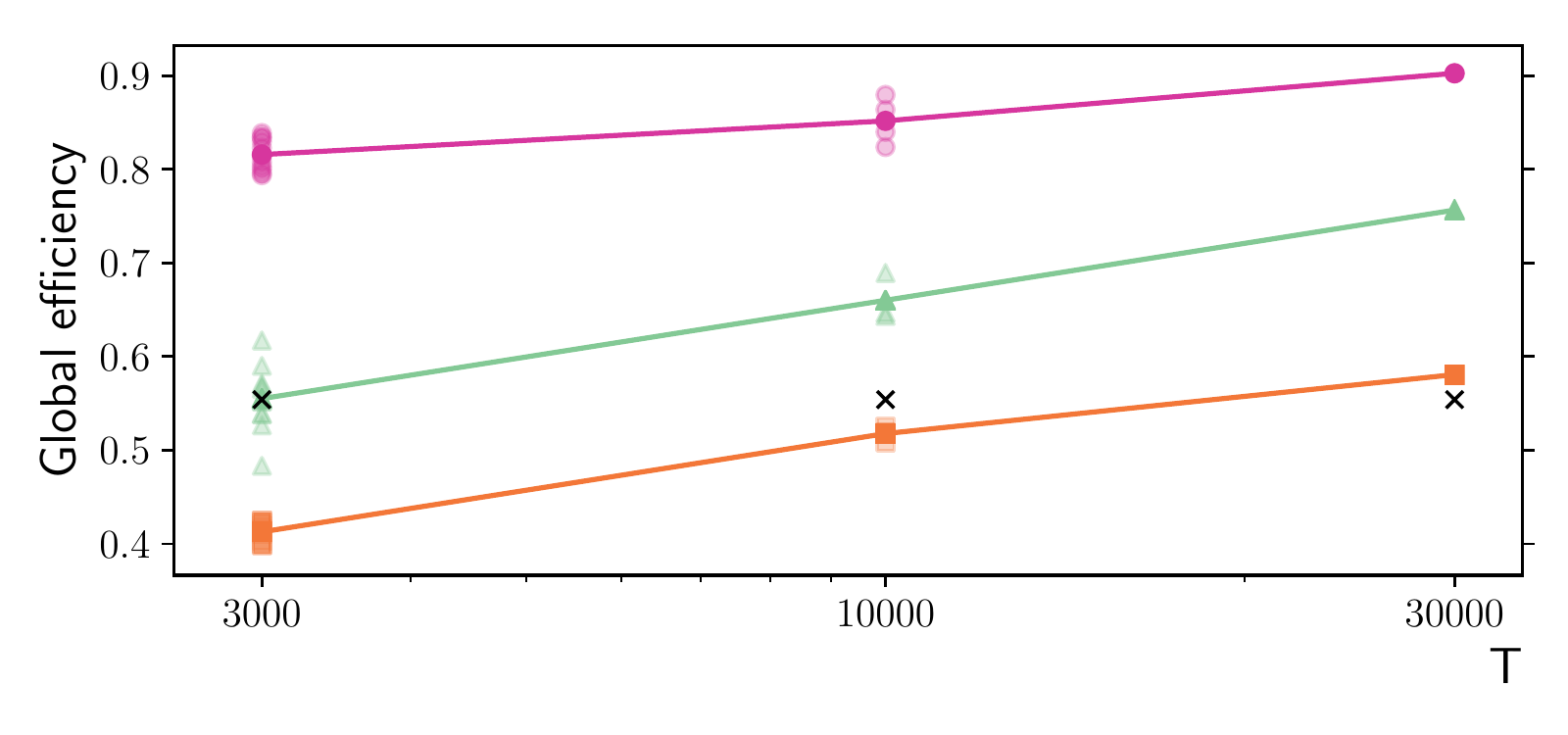}
    \fi
   
    \ifarXiv\includegraphics[width=0.45\textwidth]{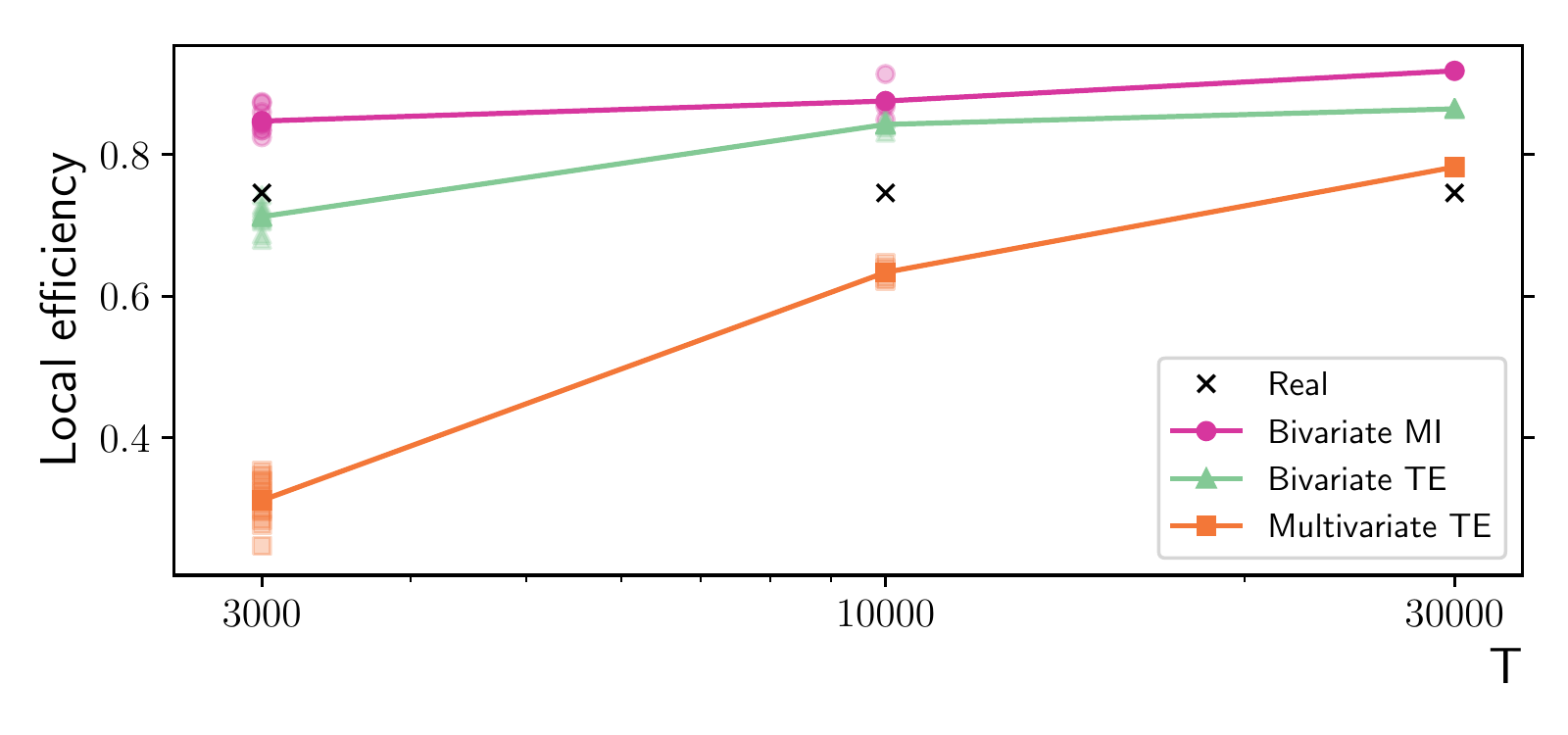}
    \fi
    \caption{\label{fig:macaque76_efficiency_vs_T}
        Global and local efficiency as a function of number of time samples $T$ in a real macacque connectome with $N$=\num{76} nodes.
        Bivariate MI consistently overestimates both measures.
        Bivariate TE produce the most accurate estimates for shorter time series but tends to overestimate both efficiency measures as more data is provided.
        Multivariate TE significantly underestimates both measures for shorter time series, but is the only method that approximately converges to real values as more data is provided.
        The results for all simulations are presented (low-opacity markers) in addition to the mean values (solid markers).
    }
\end{figure}

\ifarXiv
\section{Local and global efficiency}
\fi
\label{app:efficiency}
The crucial limitation of shortest path length measures (and of the derived small-world coefficient) is being only defined for connected networks.
Therefore, the analogous global efficiency measure~\citep{Latora2001} is often used to overcome this shortcoming (also see~\citep{Zanin2015} for an alternative measure of small-worldness based on global efficiency).
The related local efficiency measure can instead be regarded as analogous to the clustering coefficient~\citep{Latora2001}.
Complementing the results in the main text, we report the global and local efficiency of small-world networks (\fig{WS_efficiency_vs_rewiring_T10000}), scale-free networks (\fig{BA_m2_efficiency}), and two real macaque connectomes (\fig{macaque71_efficiency_vs_total_cross_coupling} and \fig{macaque76_efficiency_vs_T}).

\begin{figure}
    \ifarXiv\includegraphics[width=0.48\textwidth]{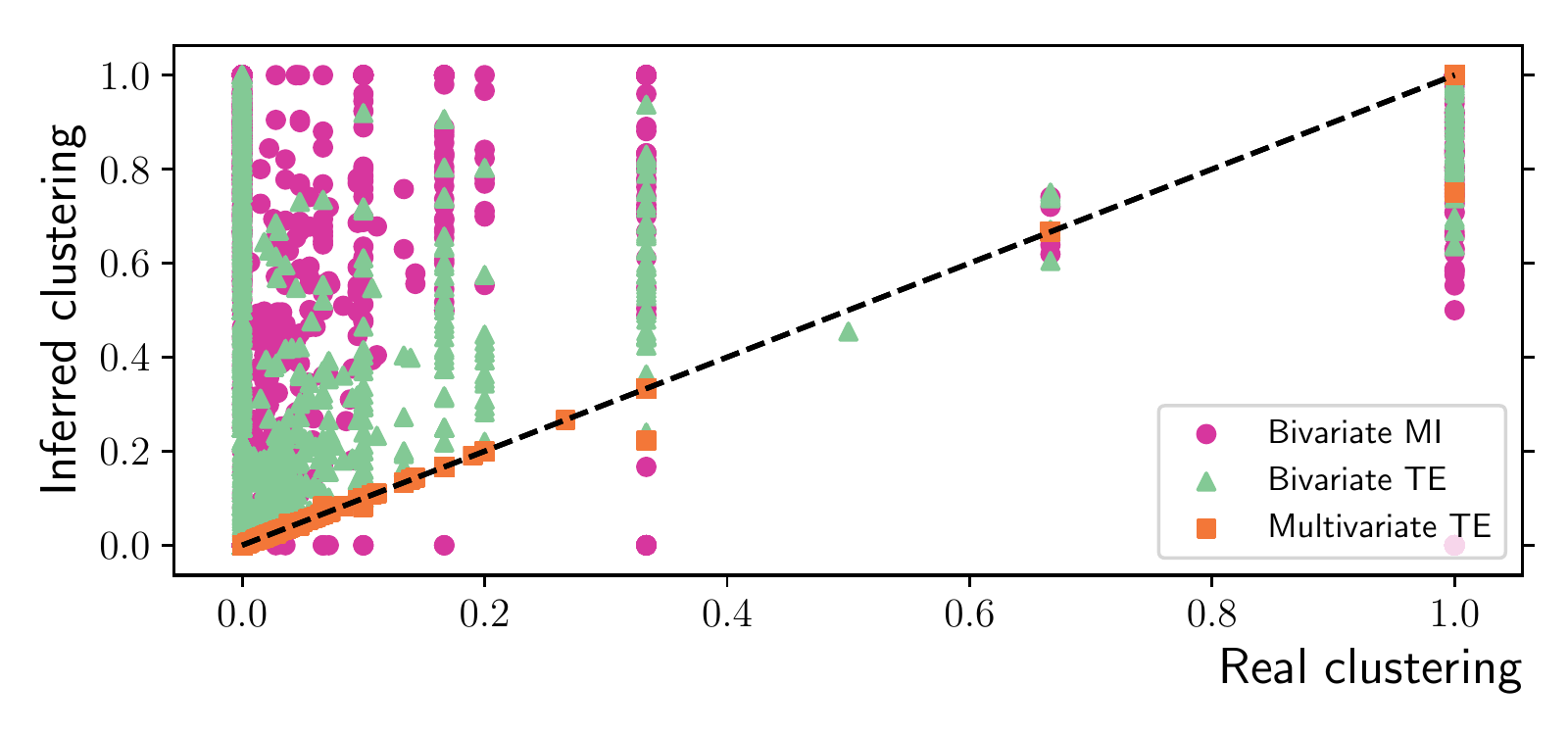}
    \fi
    \caption{\label{fig:BA_m2_clustering}
        Inferred vs. real clustering coefficient of individual nodes in scale-free networks obtained via preferential attachment ($N$=\num{200} nodes and $T$=\num{10000} time samples).
        Multivariate TE accurately reproduces the clustering coefficient of the real networks (ground truth), while bivariate methods overestimate low clustering values and underestimate high clustering values.
        The results are collected over~\num{10} simulations on different network realisations.
    }
\end{figure}

\ifarXiv
\section{Clustering coefficient of scale-free networks}
\fi
\label{app:BA_clustering}
Plotting the clustering coefficient values of individual nodes instead of the average shows that the low clustering values are consistently overestimated by bivariate methods (which is the most prominent effect affecting the average), while high clustering values are underestimated (\fig{BA_m2_clustering}).

\ifarXiv
\section{Rich-club coefficient and assortativity of scale-free networks}
\fi
\label{app:BA_rich_club}

\begin{figure}
    \ifarXiv\includegraphics[width=0.48\textwidth]{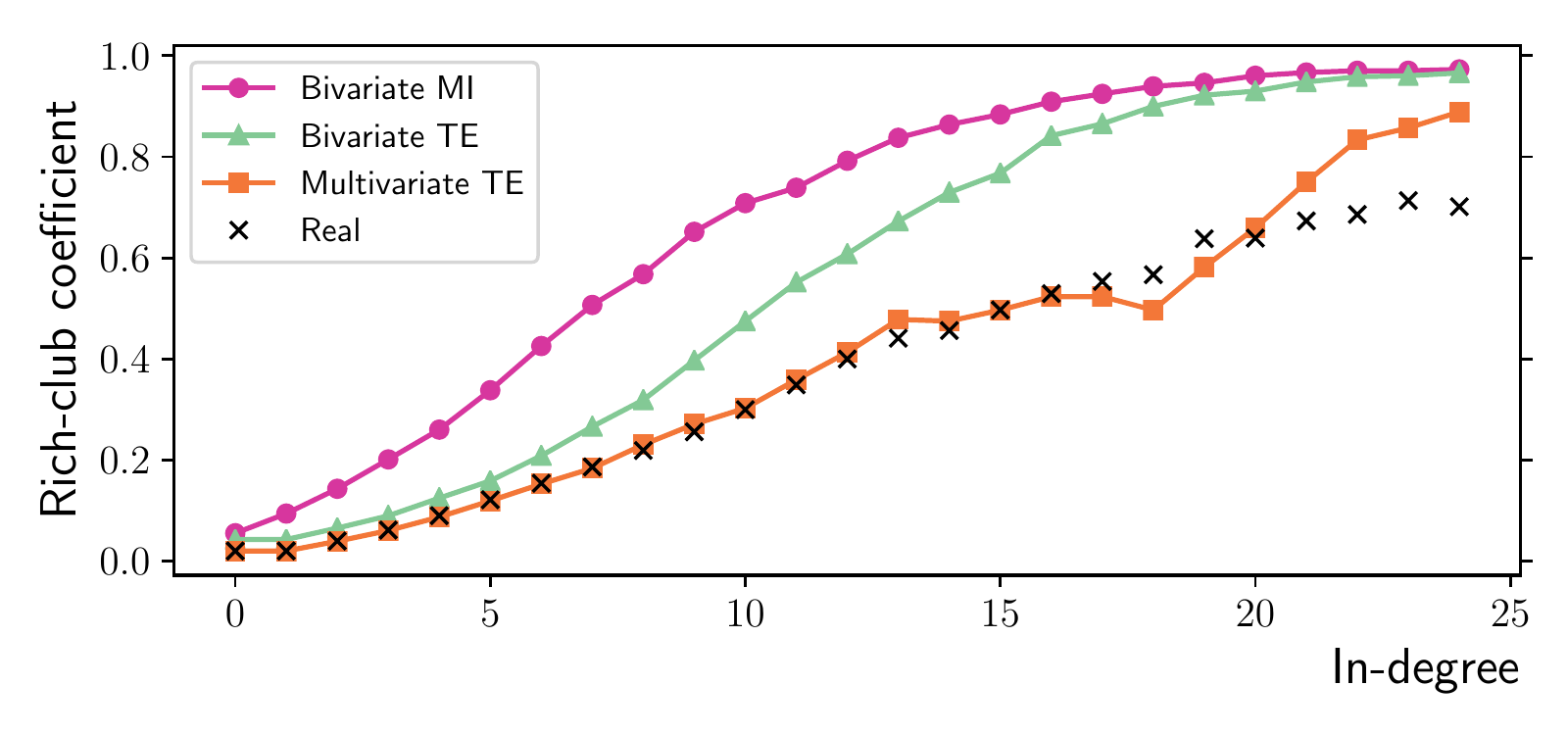}
    \fi
    \caption{\label{fig:BA_m2_rich_club_in_degree}
        Rich-club coefficient in scale-free networks obtained via preferential attachment ($N$=\num{200} nodes and $T$=\num{10000} time samples).
        Multivariate TE accurately reproduces the rich-club coefficient of the real networks (ground truth) for lower in-degree thresholds, while bivariate TE and MI consistently overestimate it.
        Mean values over~\num{10} simulations.
    }
\end{figure}

The rich-club coefficient measures the extent to which high-degree nodes connect to each other~\citep{Colizza2006}.
Instead of choosing a specific threshold to define high-degree nodes, the rich-club coefficient is plotted in \fig{BA_m2_rich_club_in_degree} for a range of thresholds (non-normalised values).
The rich-club coefficient is overestimated by bivariate MI and TE across all thresholds, although the effect is less prominent than on other network properties. 

\begin{figure}
    \ifarXiv\includegraphics[width=0.48\textwidth]{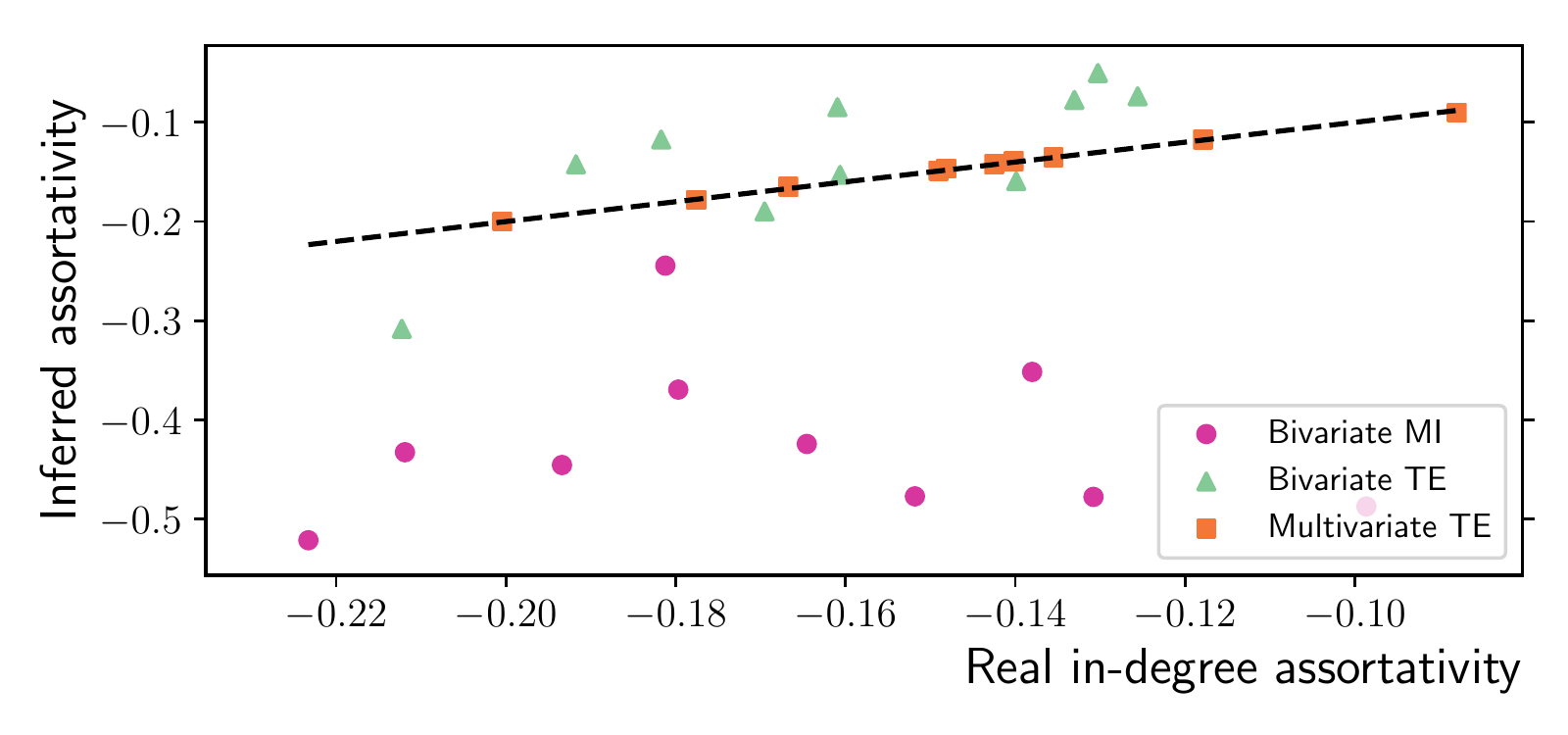}
    \fi
    \caption{\label{fig:BA_m2_assortativity_in_degree}
        In-degree assortativity coefficient in scale-free networks obtained via preferential attachment ($N$=\num{200} nodes and $T$=\num{10000} time samples).
        Bivariate and multivariate TE accurately reproduce the assortativity of the real networks (ground truth), while bivariate MI consistently underestimate it.
        The black dashed line represents the identity between real and inferred values.
        The results are shown for \num{10} different network realisations.
    }
\end{figure}

Assortativity (or assortative mixing) is a preference for nodes to attach to others with similar degree.
The in-degree assortativity coefficient is the Pearson correlation coefficient of degree between pairs of linked nodes.
Positive values indicate a correlation between nodes of similar in-degree, while negative values indicate relationships between nodes of different in-degree.
As shown in \fig{BA_m2_assortativity_in_degree}, the scale-free networks obtained via preferential attachment are disassortative (\ie they have negative assortativity coefficients).
Bivariate and multivariate TE accurately reproduce the assortativity of the real networks (ground truth), while bivariate MI consistently underestimate it.

\ifarXiv
\section{Reducing $\alpha$ and sample ROC curve}
\fi
\label{app:roc_curve}

Reducing the critical statistical significance level $\alpha$ clearly helps to reduce false positives, for any approach.
However, what distinguishes the multivariate approach on this point is two-fold.

First, the significance level $\alpha$ has direct meaning regarding the false-positive rate (FPR) under idealised conditions, implying that a well-calibrated test should produce a FPR consistent with $\alpha$.
This is the case for multivariate TE under the ideal conditions investigated here, but not for bivariate measures, where the FPR is inflated drastically above the requested level.
Certainly one can decrease $\alpha$, but the experimenter has no a-priori insights regarding what to set it to.

Second, even though one can in principle decrease the FPR for bivariate measures by decreasing the significance level, a very large number of true positives would also be missed by doing so, and therefore the desired reflection of the relationships in the underlying structure would not be achieved.

\begin{figure}
    \ifarXiv\includegraphics[width=0.48\textwidth]{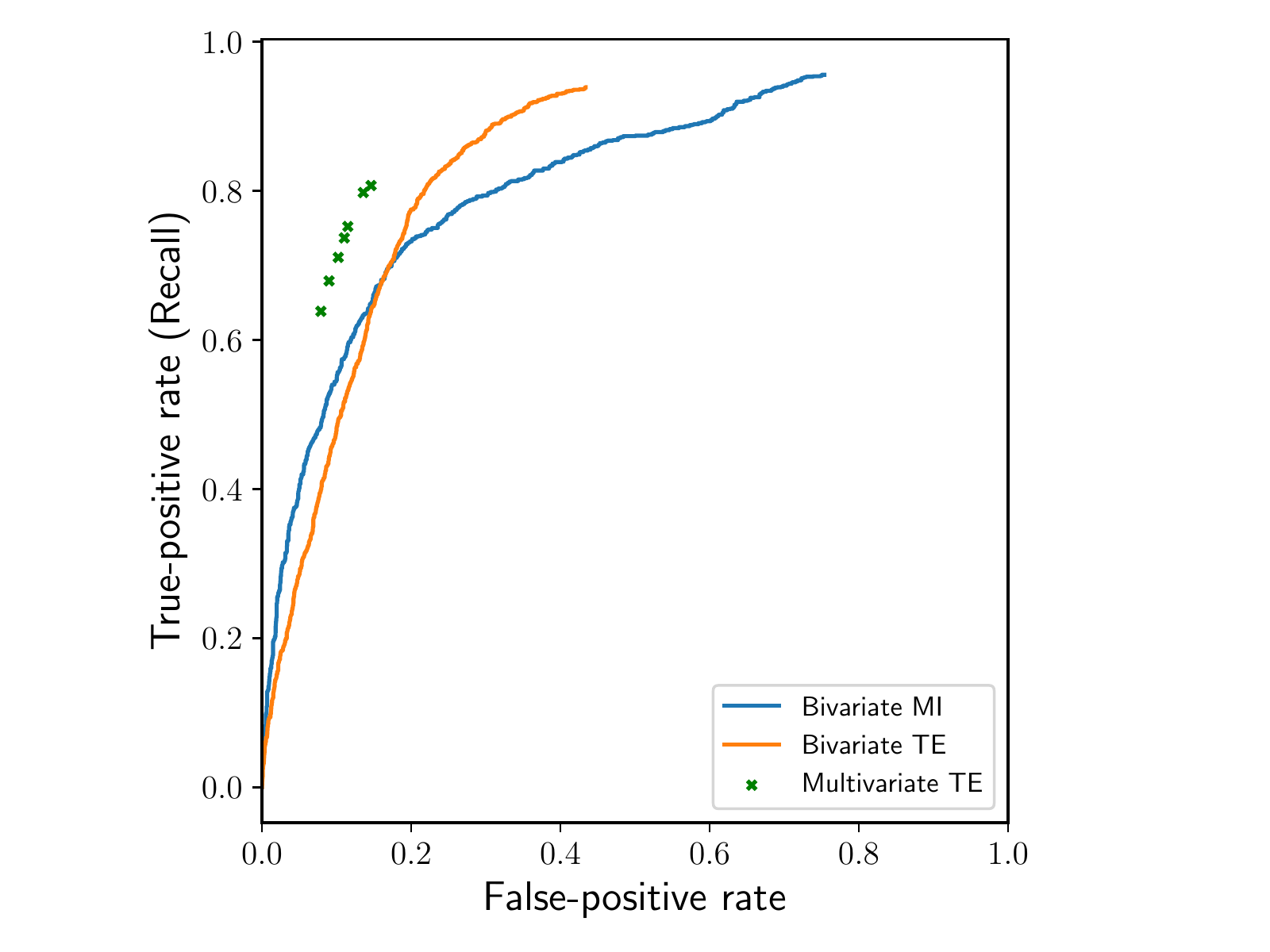}
    \fi
    \caption{\label{fig:macaque76_30000samples_ROC}
        Partial receiver operating characteristic curve. For multivariate TE, note that the whole algorithm must be re-run for each $\alpha$, and so a limited locus of the ROC curve is plotted for it. The specific threshold values tested for multivariate TE are $\alpha~\in~\{0.00001, 0.0001, 0.001, 0.005, 0.01, 0.05, 0.1\}$.
    }
\end{figure}

We can compare the receiver operating characteristic (ROC) curve for bivariate measures to the multivariate TE for the experiment on a neural mass model, shown in \fig{macaque76_30000samples_ROC}.
Note that for multivariate TE, each point on the scatter plot is generated from separate runs with different $\alpha$ parameters rather than a single run (since the recall and FPR are functions of the whole inferred network).
We see that the FPRs in this experiment are substantially inflated over the experiments under idealised conditions, as discussed in the main text.
Crucially though, whilst these are inflated for all measures, the contrasts in FPRs between the approaches are quite large when converted to the numbers of spurious links inferred.
This can be seen visually in \fig{macaque76_30000samples_adj_mats}.
The multivariate TE operates at a much larger true-positive rate than the bivariate measures for the same FPR; therefore, simply reducing $\alpha$ for bivariate measures is not going to result in as effective a performance as multivariate TE.

\clearpage 
\bibliography{bibliography}

\end{document}